\documentclass[12pt]{article}
\pdfoutput=1
\usepackage{amsmath,amsfonts,graphicx,color,bbm,tikz,subcaption}
\usepackage{comment}
\usepackage[nosort]{cite}
\usetikzlibrary{calc,positioning}
\usetikzlibrary{patterns,arrows,decorations.pathreplacing}
\usepackage{caption}
\usepackage{ulem}
\tikzset{>=stealth}
\usepackage{lipsum}
\usepackage{tabularx}
\usepackage{fullpage}
\usepackage{hyperref}
\usepackage{color,soul}

\definecolor{red}{RGB}{ 211, 0, 0}
\definecolor{blue}{RGB}{1, 62, 177}

\makeatletter\@addtoreset{equation}{section}\makeatother

\newcommand{\be}{\begin{equation}}
\newcommand{\ee}{\end{equation}}
\def\beq{\begin{equation}}
\def\eeq{\end{equation}}
\newcommand{\bea}{\begin{eqnarray}}
\newcommand{\eea}{\end{eqnarray}}

\newcommand{\vev}[1]{{\left< {#1} \right>}}

\newcommand{\ket}[1]{{\left| {#1} \right>}}

\renewcommand{\title}[1]{\vbox{\center\LARGE{#1}}\vspace{3mm}}
\renewcommand{\author}[1]{\vbox{\center{#1}}\vspace{3mm}}

\newcommand{\email}[1]{\vbox{\center\tt#1}\vspace{3mm}}

\begin{document}
\begin{titlepage}

\begin{center}
{\large {\bf Disorder-free localisation in continuous-time quantum walks \\ - Role of symmetries -}}

\author{A. P. Balachandran,$^{a}$ Anjali Kundalpady,$^b$
Pramod Padmanabhan,$^c$ Akash Sinha$^c$ }

{$^a${\it Physics Department, Syracuse University, \\ Syracuse, NY, 13244-1130, USA}}
\vskip0.1cm
{$^b${\it The International Centre for Theoretical Sciences (ICTS),\\  Survey No. 151, Shivakote, Hesaraghatta Hobli, Bengaluru, 560089, India}}
\vskip0.1cm
{$^c${\it School of Basic Sciences,\\ Indian Institute of Technology, Bhubaneswar, 752050, India}}

\email{apbal1938@gmail.com, anjali.kundalpady@gmail.com, pramod23phys@gmail.com,  akash.sinha@gmail.com}

\vskip 0.5cm 

\end{center}


\abstract{
\noindent 
We investigate the phenomenon of disorder-free localisation in quantum systems with global permutation symmetry. We use permutation group theory to systematically construct permutation symmetric many-fermion Hamiltonians and interpret them as generators of continuous-time quantum walks. When the number of fermions is very large we find that all the canonical basis states localise at all times, without the introduction of any disorder coefficients. This time-independent localisation is not the result of any emergent disorder distinguishing it from existing mechanisms for disorder-free localisation. Next we establish the conditions under which the localisation is preserved. We find that interactions that preserve and break the global permutation symmetry sustains localisation. Furthermore the basis states of systems with reduced permutation symmetry, localise even for a small number of fermions when the symmetry-reducing parameters are tuned accordingly. 
We show that similar localisation also occurs for a permutation symmetric Heisenberg spin chain and permutation symmetric bosonic systems, implying that the localisation is independent of the superselected symmetry. Finally we make connections of the Hamiltonians studied here to the adjacency matrices of graphs and use this to propose a prescription for disorder-free localisation in continuous-time quantum walk systems. Many of the models proposed here feature all-to-all connectivity and can be potentially realised on superconducting quantum circuits, trapped ion systems and ultracold atoms.}

\end{titlepage}

\section{Introduction}
\label{sec:Introduction}
A closed many-body system is described by a few parameters (temperature, pressure etc.) once it attains equilibrium. This thermal behavior, while true in non-integrable classical systems, may not always hold for their quantum analogues where the effects of destructive interference among the constituents can result in {\it localised states} that are non-thermal \cite{1958PhRv..109.1492A, Basko2005MetalinsulatorTI} thus violating the {\it eigenstate thermalisation hypothesis} (ETH) \cite{PhysRevE.50.888, PhysRevA.43.2046, Deutsch_2018, Buvca2023UnifiedTO}. The typical mechanism to produce such states is to use random potentials or introduce disorder coefficients in the Hamiltonian. However there exist methods to produce localised states without disorder, now known as {\it disorder-free localisation} (DFL), showing that disorder is just a sufficient feature and not necessary. Systems showing this phenomena include lattice gauge theories \cite{PhysRevLett.118.266601, PhysRevLett.119.176601, PhysRevLett.120.030601, PhysRevB.102.165132} and the so called Stark MBL models \cite{PhysRevLett.122.040606, Morong2021PublisherCO, PhysRevB.106.174305, Wang2021StarkML} among others \cite{Gao2023NonthermalDI, PhysRevResearch.3.L032069, Wadleigh2022InteractingSL, Zhang2022StableIA}. In much of these systems the disorder is emergent in the Hilbert space dynamics resulting in the ergodicity-breaking property. 

In this work we show a completely different origin of DFL in systems with and without {\it superselected} and global symmetries. Our method is to use the properties of the spectrum of Hamiltonians to produce localisation of states in Hilbert space. We observe that systems where one eigenvalue has a large degeneracy compared to others can result in Hilbert space localisation without any disorder. We construct such Hamiltonians on the Fock space and find that all the basis states of each number sector localise. Before we proceed further we clarify the definition of localisation we will use. Consider a state $\ket{\psi}$, a basis element of some number sector of the Fock space, evolving under a Hamiltonian $H$. Then if the probability distribution computed from the overlap between $$\ket{\psi(t)}=e^{-\mathrm{i}Ht}\ket{\psi(0)}$$ and $\ket{\psi(0)}$ stays close to one at all times, we say that the state $\ket{\psi}$ is localised\footnote{This quantity is also related to the Loschmidt echo \cite{Peres1984StabilityOQ, Goussev2012LoschmidtE}.}\footnote{In this definition, eigenstates of the system will be localised. For the systems we study we will consider localised states that does not include the eigenstates.}. In other words the time evolved states in this system retain memory of the initial states at all times, $t$. This definition implies that the states of our systems localise in Hilbert space. And due to this there is no notion of localisation lengths for our states and in this regard it is similar to localisation {\it via} MBL states. \cite{Abanin2018ColloquiumM}. 

The systems we will consider include various symmetries including {\it superselected symmetries}. Typically a quantum theory with {\it superselection sectors} is tightly constrained, restricting the possible representations that the algebra of observables can take. The operators of such theories commute with the superselected symmetries and the Hilbert space splits into different superselection sectors that are labelled by the irreducible representations of the superselected symmetries. A physically important example of such a symmetry is the {\it permutation symmetry} generated by the statistics operator that occurs while studying a system of identical and indistinguishable {\it bosons} ({\it fermions}). In such theories the states are symmetrised (antisymmetrised) and the operators are invariant under this exchange symmetry. 

In general such systems, though constrained, are not expected to show non-thermal behaviour. For example the {\it tight-binding Hamiltonian} of fermions or bosons do not show any localisation without disorder. In this work we will exhibit several fermionic Hamiltonians with and without global symmetries that show DFL. To construct these models we begin with a many-fermion Hamiltonian with global permutation symmetry $\mathcal{S}_N$. This Hamiltonian is number-preserving and we find that all the canonical basis states in each fermion number sector localises without any disorder for large $N$. The requirement of the global $\mathcal{S}_N$ symmetry enforces the interaction of each fermion with every other fermion making the Hamiltonian similar in appearance to the SYK model \cite{KitaevSeminar,Sachdev1992GaplessSG,Maldacena2015ABO,Rosenhaus2018AnIT}. 
Our naive intuition might suggest that such a system where all particles interact with each other should thermalise. On the contrary we see complete localisation in their dynamics. It is also worth noting that there is no emergent disorder in the models described here unlike the gauged models in the literature that exhibit DFL. 
Our construction also considers permutation symmetric operators in higher-fermion number sectors which can be thought of as interactions. The localisation is stable to the inclusion of such terms. Additionally there are terms that break the global $\mathcal{S}_N$ symmetry but preserves the localisation. We show several examples of such terms which both preserve and destroy the localisation. 

Furthermore the models we write down can be interpreted as quantum walk Hamiltonians that have received a lot of attention in the past \cite{Kempe2003QuantumRW,VenegasAndraca2012QuantumWA, Reitzner2012QuantumW, Childs2001AnEO, Ambainis2001OnedimensionalQW, aharonov1993quantum} in the context of search algorithms \cite{Santha2008QuantumWB, Shenvi2002QuantumRS, 10.5555/2462630}, as quantum simulators \cite{Bepari2021QuantumWA,Nachman2019QuantumAF,Somma2008QuantumSO, Strauch2005RelativisticQW} and for universal quantum computation \cite{Childs2012UniversalCB,Underwood2010UniversalQC,Asaka2021TwolevelQW, Childs2008UniversalCB}. We are concerned with {\it continuous-time quantum walks} (CTQW) of identical particles \cite{Sansoni2011TwoparticleBQ, Qin2014StatisticsdependentQC,Giri2020TwoCQ,Melnikov2013QuantumWO,Melnikov2018HittingTF,Lahini2011QuantumWO,Wiater2017TwoBQ} and especially those with an additional global symmetry. Symmetric quantum walks have been considered in the case of {\it discrete-time quantum walks} (DTQW) \cite{Krovi2007SymmetryIQ, Janmark2014GlobalSI, PhysRevA.76.022316} and have been shown to feature topological phases \cite{Cedzich2016TheTC, Cedzich2019QuantumWS, Geib2019TopologicalAO, Cedzich2018CompleteHI} and localisation \cite{Danac2020DisorderfreeLI,Mandal2021LocalizationOT, Singh2017InterferenceAC, Joye2012DynamicalLF, Cedzich2019AndersonLF}. 

The rest of the article is organised as follows. We begin Sec. \ref{sec:construction} with the space describing the fermions and the operators acting on them. The Hamiltonians with global $\mathcal{S}_N$ symmetry are constructed using simple ideas from the theory of permutation groups. Among the many possibilities, we consider the simplest such Hamiltonian in Sec. \ref{subsec:2-cycleH}. The solution of the Hamiltonian is provided in Sec. \ref{sec:solution} and the resulting features are compared with the models that lack a global $\mathcal{S}_N$ symmetry. 

In a rather long discussion in section \ref{sec:discussion}, we explore several other models that both display and break the DFL discussed for the $\mathcal{S}_N$-symmetric model in the first part of the paper. In particular we study the effect of perturbations that break the global symmetry in Sec. \ref{subsec:perturbations}. Next we study when the localisation persists for different initial conditions in Sec. \ref{subsec:initialconditions}. In Sec. \ref{subsec:roleofsymmetries} we analyse the role played by the global symmetry in the localisation and find that a fermionic model with a reduced global symmetry ($\mathcal{S}_N\rightarrow\mathcal{S}_{N-k}$) can also exhibit localisation. We note here that a spin system with a global $\mathcal{S}_N$ symmetry and a permutation symmetric bosonic system also display similar localisation features. Then we highlight an interesting connection of these systems to adjacency matrices of graphs in Sec. \ref{subsec:DFLgraphs}. This connection suggests a possible way to generalise the DFL studied in this work.
We then conclude with a few remarks about experimental realisations and future theoretical directions in the same section.

\section{Construction}
\label{sec:construction}
We begin with a brief description of the Fock space (spanned by the states diagonalising the number operator) describing $N$ identical and indistinguishable fermions. The {\it vacuum}, $\ket{\Omega}$ denotes the state with no fermions. The state $\ket{i}$, for $i\in\{1,2,\cdots, N\}$, describes the presence of a fermion on site $i$. These are the 1-fermion states and they span a $N$ dimensional space henceforth denoted $\mathcal{H}\simeq\mathbb{C}^N$ with the canonical inner product. 
In this notation, multiparticle states such as, $\ket{i}\otimes\ket{j}$ live in $\mathcal{H}\otimes\mathcal{H}$. No relation between $\ket{i}\otimes\ket{j}$ and $\ket{j}\otimes\ket{i}$ is assumed a priori. However in the case of indistinguishable fermions, we work with normalised antisymmetrised states, $\frac{1}{\sqrt{2}}\left[\ket{i}\otimes\ket{j} - \ket{j}\otimes\ket{i} \right]$ that live in $\mathcal{H}\wedge\mathcal{H}$,
with the $\wedge$ denoting antisymmetrisation. Thus the full Hilbert space becomes the antisymmetrised Fock space,
$$ \bigoplus\limits_{n=0}^N~\mathcal{H}^{\wedge n}.$$
This space is finite and its dimension is seen from
$$ \sum\limits_{k=0}^N~\left(\begin{array}{c} N \\ k \end{array}\right) = 2^N,$$
with $\left(\begin{array}{c} N \\ k \end{array}\right)$ being the dimension of $\underbrace{\mathcal{H}\wedge\mathcal{H}\wedge\cdots\wedge\mathcal{H}}_{k~\textrm{times}}$.
  
The creation ($a_j^\dag$) and annihilation ($a_j$) operators satisfying the fermionic (CAR) algebra,
\begin{eqnarray}\label{eq:CAR}
\{ a_j, a_k^\dag\} & = & \delta_{jk}, \nonumber \\  
\{ a_j, a_k\} & = & \{ a_j^\dag, a_k^\dag\} = 0,
\end{eqnarray}
(where $\{a,b\}:= a b + b a$) are realized on this space. The index $j$ take values in, $\{1,\cdots, N\}$. 
More generally we could add an extra index $\mu$ to each oscillator to denote an internal degree of freedom like a {\it colour} or {\it spin} index. For simplicity we will stick to just the indices $j\in\{1,2,\cdots, N\}$ for the fermions, allowing their interpretation as lattice sites.

An arbitrary $k$-fermion state expressed as 
\begin{eqnarray}
a_{i_1}^{\dagger}a_{i_2}^{\dagger}\cdots a_{i_k}^{\dagger}|\Omega\rangle,
\end{eqnarray}
 satisfies all the necessary antisymmetry properties as can be directly verified by using \eqref{eq:CAR}. With the help of these $a$ and $a^{\dagger}$, we can mutate between different particle sectors. For example, the action of $a_i^{\dagger}$ on an arbitrary state from $ \mathcal{H}^{\wedge k}$ yields 
 \begin{eqnarray}
 a_i^{\dagger}\left(a_{i_1}^{\dagger}a_{i_2}^{\dagger}\cdots a_{i_k}^{\dagger}|\Omega\rangle\right)=a_i^{\dagger}a_{i_1}^{\dagger}a_{i_2}^{\dagger}\cdots a_{i_k}^{\dagger}|\Omega\rangle\in \mathcal{H}^{\wedge k+1}.
\end{eqnarray}
The Fock space description ensures that the fermionic creation and annihilation operators commute with the superselected exchange symmetry of this system. 

Next we move on to the action of the global permutation symmetry $\mathcal{S}_N$ on the site indices $\{1,2,\cdots, N\}$. The action of these operators on the states and operators of the theory are obtained as follows.
The vacuum is invariant under permutations, $s_{ij}\ket{\Omega} = \ket{\Omega}$ and the transformation rule of operators under conjugation by permutation generators is, $$s_{jk}O_{\cdots j\cdots k\cdots }s_{jk}^{-1}=O_{\cdots k\cdots j\cdots},$$
where $O$ is an operator with several indices including $j$ and $k$ (Note that $s_{jk}^{-1}=s_{jk})$. Using these properties we can deduce the action of the permutation group on arbitrary states of this system. 

Permutation invariant operators acting on these states satisfy 
\begin{equation}
    s_iO_{i_1\cdots i_p}s_i = O_{i_1\cdots i_p},~~\forall~i\in\{1,2,\cdots, N-1\},
\end{equation}
where $s_i\equiv s_{i, i+1}$, are the transposition operators that generate the permutation group $(\mathcal{S}_N)$ and satisfy
\begin{equation}\label{eq:SNrelations}
s_is_{i+1}s_i=s_{i+1}s_is_{i+1},~~s_i^2=1,~~s_is_j=s_js_i~\textrm{when}~|i-j|\geq 2.
\end{equation}

For the particular value of $N=2$, let us consider the two operators
\begin{eqnarray}
a_1^{\dagger}a_1+a_2^{\dagger}a_2\quad\quad\quad\text{and}\quad\quad\quad a_1^{\dagger}a_2+a_2^{\dagger}a_1.
\end{eqnarray}
They clearly are invariant under the action of $\mathcal{S}_2$ and the objective is to construct such permutation invariant operators for arbitrary $\mathcal{S}_N$. A natural place to look for such objects are in the conjugacy classes of $\mathcal{S}_N$ which are left invariant as a set under the action of the group by definition. For the permutation group, the elements of a conjugacy class have the same {\it cycle} structure and their order is given by $$\frac{N !}{\prod\limits_{k=1}^N~\left(k^{\nu_k}\right)\nu_k ! }$$ where $\nu_k$ is the number of $k$-cycles. A clear invariant is the sum of the elements of a given conjugacy class with a particular cycle structure\footnote{These are precisely the generators of the center of the permutation group algebra $\mathbb{C}(\mathcal{S}_N)$.}. These statements are independent of the particular realisation of the transposition operators. For our current problem we will show a realisation using the fermionic creation and annihilation operators. The operators we use are such that the resulting Hamiltonians are hermitian and number-preserving as the permutation operators do not change the number of fermions. 

We will obtain the fermionic realisation of $\mathcal{S}_N$ by showing the existence of the permutation operators for each particle sector. The fermionic realisation for a generic transposition permutation $\sigma\in\mathcal{S}_N$ in the $k$-fermion sector is given by,
\begin{equation}\label{eq:fpermutation-kfermion}
    \sigma = \sum\limits_{i_1<i_2<\cdots i_k}~a^\dag_{\sigma(i_1)} a^\dag_{\sigma(i_2)}\cdots a^\dag_{\sigma(i_k)}a_{i_k}\cdots a_{i_2}a_{i_1}.
\end{equation}
The permutation $\sigma$ has a particular cycle structure. For example the fermionic realisations of the transpositions $(ij)$ in the 1-fermion and 2-fermion sectors are given by
\begin{equation}\label{eq:ftransposition-1fermion}
    (ij)_1 = a^\dag_ja_i + a^\dag_ia_j + \mathop{\sum_{k=1}^N}_{k\neq i,j}~a^\dag_k a_k, 
\end{equation}
and 
\begin{equation}\label{eq:ftransposition-2fermion}
    (ij)_2 = a^\dag_ja^\dag_ia_ja_i + \sum\limits_{k\neq j}~a^\dag_ja^\dag_ka_ka_i + \sum\limits_{k\neq i}~a^\dag_ka^\dag_ia_ja_k + \mathop{\sum_{k,l=1}^N}_{k<l\neq\{i,j\}}~a^\dag_ka^\dag_la_la_k,
\end{equation}
respectively. The suffix $\alpha$, on $(ij)_\alpha$ denotes fermion number sector on which this transposition acts. Thus on the full Fock space the transposition is given by,
\begin{equation}\label{eq:ftransposition-fullspace}
    (ij) = \bigoplus\limits_{\alpha=1}^N~(ij)_\alpha.
\end{equation}
A more non-trivial example is that of a 3-cycle permutation in the 1-fermion sector,
\begin{equation}\label{f3cycle-1fermion}
    (ijk)_1 = a^\dag_ja_i + a^\dag_ka_j + a^\dag_ia_k + \mathop{\sum_{l=1}^N}_{l\neq\{i,j,k\}}~a^\dag_la_l.
\end{equation}
Note that the hermitian conjugate of this term is $(ikj)_1$. Indeed the Hamiltonian identified as a sum of the elements of the conjugacy class will turn out to be hermitian. We can now write down an $\mathcal{S}_N$ invariant Hamiltonian for a given conjugacy class made of $p$-cycles as
\begin{equation}\label{eq:Hgeneralconjugacyclass}
    H^{(p)} = \bigoplus\limits_{\alpha=1}^N~H^{(p)}_\alpha,
\end{equation}
where 
\begin{equation}\label{eq:Hpalpha}
    H^{(p)}_\alpha = \sum\limits_{i_1<i_2<\cdots <i_p}~(i_1i_2\cdots i_p)_\alpha,
\end{equation}
acts on the $\alpha$-fermion sector. Such $\mathcal{S}_N$ invariant Hamiltonians \eqref{eq:Hgeneralconjugacyclass} are true for any realisation of the permutation group. The fermionic realisation in \eqref{eq:fpermutation-kfermion} introduces simplifications to the Hamiltonian due to the CAR algebra \eqref{eq:CAR}. 

Before going into these we first note that the operators corresponding to cycles of length larger than $\beta$ annihilate the vectors in the $\beta$-fermionic sector. Thus the bilinear expression acting on the 1-fermion sector affects all possible fermion number sectors in a system of $N$ fermions. It acts as an exchange operator on the 1-fermion states and has a non-trivial action on the remaining sectors.

In what follows we will restrict ourselves to the Hamiltonians constructed out of the 2-cycle conjugacy class. We will comment on models obtained from other conjugacy classes in Sec. \ref{sec:discussion} and carry out a more detailed investigation in a future work.

\subsection{2-cycle Hamiltonian}
\label{subsec:2-cycleH}
Consider the conjugacy class made out of purely 2-cycles which are just the transpositions. They include the exchange of any two of the $N$ indices and there are precisely $\frac{N(N-1)}{2}$ of them. The 2-cycle Hamiltonian in the 1-fermion sector is obtained using \eqref{eq:ftransposition-1fermion} and is bilinear in the fermion creation and annihilation operators,  
\begin{equation}\label{eq:2-cycleH-1fermion}
    H^{(2)}_1 = \sum\limits_{i<j}~ \left[a_i^\dag a_j + a_j^\dag a_i\right] + \frac{(N-1)(N-2)}{2}~\hat{N}.
\end{equation}
 The factor accompanying the number operator $\hat{N}=\sum\limits_{k=1}^N~a^\dag_ka_k$ is a result of the substitution \eqref{eq:ftransposition-1fermion} for the 2-cycles. Clearly the term in the $\left[\cdots\right]$ commutes with $\hat{N}$ and represents a fermion on a given site hopping to any other site. As noted earlier this Hamiltonian has a non-trivial action on every fermion number sector except on the 1-fermion sector where it acts as a permutation operator. It is clearly $\mathcal{S}_N$ invariant in its site indices\footnote{A more rigorous proof is shown in App. \ref{app:H1 every sector}.}. The second term in \eqref{eq:2-cycleH-1fermion} dominates for large $N$. Our goal is to study the localisation features of this system for large $N$ and so the explicit presence of $N$ in the Hamiltonian can lead to incorrect conclusions about the origin of the localisation. To avoid this we will choose the term in $\left[\cdots\right]$ as our Hamiltonian,
 \begin{equation}\label{eq:2-cycleH-alltoall}
    H = \sum\limits_{i<j}~ \left[a_i^\dag a_j + a_j^\dag a_i\right],
\end{equation}
where all the fermions interact with each other in a symmetrical manner. This model can be solved exactly by a simple change of basis as we shall see in Sec. \ref{sec:solution}.

In addition to the above bilinear Hamiltonian, we consider the operators acting on the 2-fermion states which are quartic in the fermion creation and annihilation operators. This Hamiltonian can be thought of as an interaction term when added to the bilinear Hamiltonian in \eqref{eq:2-cycleH-alltoall}. However a crucial point is that this 2-cycle Hamiltonian can be simplified\footnote{The proof for this is shown in App. \ref{app:2-cycleH-2fermion}.} using the CAR algebra in \eqref{eq:CAR} resulting in,
\begin{equation}\label{eq:2-cycleH-2fermion}
    H^{(2)}_2 = \left(\frac{(N-2)(N-3)}{2}-1 \right)\left(\hat{N}^2-\hat{N} \right) + 2\sum\limits_{i<j}~ \left[a_i^\dag a_j + a_j^\dag a_i\right]\left(\hat{N}-1 \right).
\end{equation}
This Hamiltonian continues to remain $\mathcal{S}_N$ invariant and acts on 2-fermion and higher states. These terms represent interactions but reduce to the product of bilinears due to the CAR algebra. As a consequence they commute with the Hamiltonian in \eqref{eq:2-cycleH-1fermion} and thus merely shift their eigenvalues while sharing the eigenstates. This further implies that localised states of \eqref{eq:2-cycleH-alltoall} are stable to such $\mathcal{S}_N$ preserving perturbations. This trend continues to hold for higher order perturbations obtained using the 2-cycle Hamiltonians acting on 3- and higher-fermion sectors (See App. \ref{app:2-cycleH-2fermion}).

\section{Solution}
\label{sec:solution}
The bilinear Hamiltonian in \eqref{eq:2-cycleH-alltoall} is solved with a simple change of variables in the space of creation and annihilation operators. Consider a new set of annihilation (creation) operators, $A_\alpha$ ($A_\alpha^\dag$) defined as,
\begin{equation}
    A_\alpha  =  \frac{1}{\sqrt{N}}\sum\limits_{j=1}^N~\omega^{j\alpha}~a_j,~~ 
     A_\alpha^\dag  =  \frac{1}{\sqrt{N}}\sum\limits_{j=1}^N~\omega^{-j\alpha}~a_j^\dag,
\end{equation}
with $\omega=e^{\frac{2\pi\mathrm{i}}{N}}$ being a $N$th-root of unity and $\alpha\in\{1,2,\cdots, N\}$. These operators satisfy the CAR relations required of fermionic operators,
\begin{eqnarray}\label{eq:CARalpha}
\{ A_\alpha, A_\beta^\dag\} & = & \delta_{\alpha\beta}, \nonumber \\  
\{ A_\alpha, A_\beta\} & = & \{ A_\alpha^\dag, A_\beta^\dag\} = 0.
\end{eqnarray}
In these variables, the 2-cycle Hamiltonian in \eqref{eq:2-cycleH-alltoall} reduces to 
\begin{equation}\label{eq:2-cycleHalpha}
    H = N~A_N^\dag A_N - \hat{N},
\end{equation}
where $\hat{N} = \sum\limits_{i=1}^N~a_i^\dag a_i= \sum\limits_{\alpha=1}^N~A_\alpha^\dag A_\alpha$ commutes with the Hamiltonian. A number of fermionic symmetries for the Hamiltonian in \eqref{eq:2-cycleHalpha} become apparent in this basis. We find that all bilinears $A_\alpha^\dag A_\beta$, $A_\alpha A_\beta$, $A_\alpha^\dag A_\beta^\dag$ \footnote{Removing the number operator from \eqref{eq:2-cycleHalpha} will enhance the number of fermionic symmetries as now $A_\alpha$ and $A_\alpha^\dag$ will also commute with the Hamiltonian when $\alpha\neq N$.}  commute with the Hamiltonian when $\alpha, \beta\neq N$. In fact the permutation operators of \eqref{eq:fpermutation-kfermion} can be written as linear combinations of such bilinears and thus these are the operators that map the states of a given eigenspace into each other.

The spectrum can be found by labelling the eigenspaces with the set $\{1,2,\cdots N\}$. The dimension of the $k$-fermion sector is $\frac{N\,!}{k\, !\left(N-k\right)\,!}$ and these are spanned by two kinds of eigenstates of the form
\begin{equation}\label{eq:Aalphaeigenstates}
    A_{\alpha_1}^\dag A_{\alpha_2}^\dag \cdots A_{\alpha_k}^\dag\ket{\Omega},
\end{equation}
with no two $\alpha$'s equal to each other. The first set of eigenstates are those where at least one of the $\alpha$'s is $N$. There are a total of $\frac{\left(N-1\right)\,!}{\left(k-1\right)\, !\left(N-k\right)\,!}$ such states and they share the eigenvalue $\left(N-k\right)$. The second set are those where none of the $\alpha$'s take the value $N$. These account for the remaining $\frac{\left(N-1\right)\,!}{k\, !\left(N-k-1\right)\,!}$ states and they come with the eigenvalue $-k$. In evaluating the spectrum we have used the identities,
\begin{equation}
    \left[\hat{N}, A_\alpha^\dag \right] = A_\alpha^\dag,~~ \left[\hat{N}, A_\alpha \right] = -A_\alpha. 
\end{equation}

Having obtained the spectrum, we are in a position to compute the probability distributions. We will consider 1-fermion and 2-fermion walks which sufficiently illustrate the features of the permutation invariant systems considered here. Following this we will also discuss the general $k$-fermion sector. An important point to keep in mind is the role played by the global $\mathcal{S}_N$ symmetry in determining the structure of the spectrum. For instance, it is enough to find the time evolution of any single state in a particular number sector. The remaining states can be computed by the action of the appropriate $\mathcal{S}_N$ operators on this state. Furthermore, another crucial feature arises as a consequence of the global $\mathcal{S}_N$ symmetry, namely the restriction on the subspace that a given state is allowed to evolve into. For example the 2-fermion state $\ket{1,2(t)}:=\exp(-i H t)a^{\dagger}_1a^{\dagger}_2|\Omega\rangle$ only evolves into the $\ket{1,j}$ and $\ket{2,j}$ states. There is no overlap with the states $\ket{j,k}:=a^{\dagger}_ja^{\dagger}_k|\Omega\rangle$ when $j,k\notin \{1,2\}$. In other words any state in this system does not explore the full Hilbert space under time evolution. This is not apparent from the $A_\alpha$ ($A^\dag_\alpha$) basis but becomes more transparent in a new basis. We will demonstrate this for each of the fermion number sectors below. 

\paragraph{1-fermion walks :}
The features we wish to illustrate are immediately seen in the following eigenbasis of the 1-fermion sector : there is one state of the form,
\begin{equation}\label{eq:1-fermionN-1-eigenstate}
   \sum\limits_{j=1}^N~a^\dag_j\ket{\Omega}, 
\end{equation}
with eigenvalue $N-1$ and there are $N-1$ eigenstates of the form,
\begin{equation}\label{eq:1-fermion-1-eigenstate}
   \left(a^\dag_1- a^\dag_j \right)\ket{\Omega}~;~j\in\{2,3,\cdots, N\}, 
\end{equation}
with eigenvalue $-1$. The non-degenerate state in \eqref{eq:1-fermionN-1-eigenstate} is symmetric under the action of $\mathcal{S}_N$, whereas the degenerate states in \eqref{eq:1-fermion-1-eigenstate} are mapped into each other under the action of $\mathcal{S}_N$. More precisely the transposition operators in the 1-fermion sector \eqref{eq:ftransposition-1fermion} perform this mapping. These operators can be written as linear combinations of the bilinears $A^\dag_\alpha A_\beta$ and as noted earlier these commute with the Hamiltonian \eqref{eq:2-cycleHalpha}.

\begin{figure}
\centering
\begin{subfigure}{0.49\textwidth}
    \includegraphics[width=\textwidth]{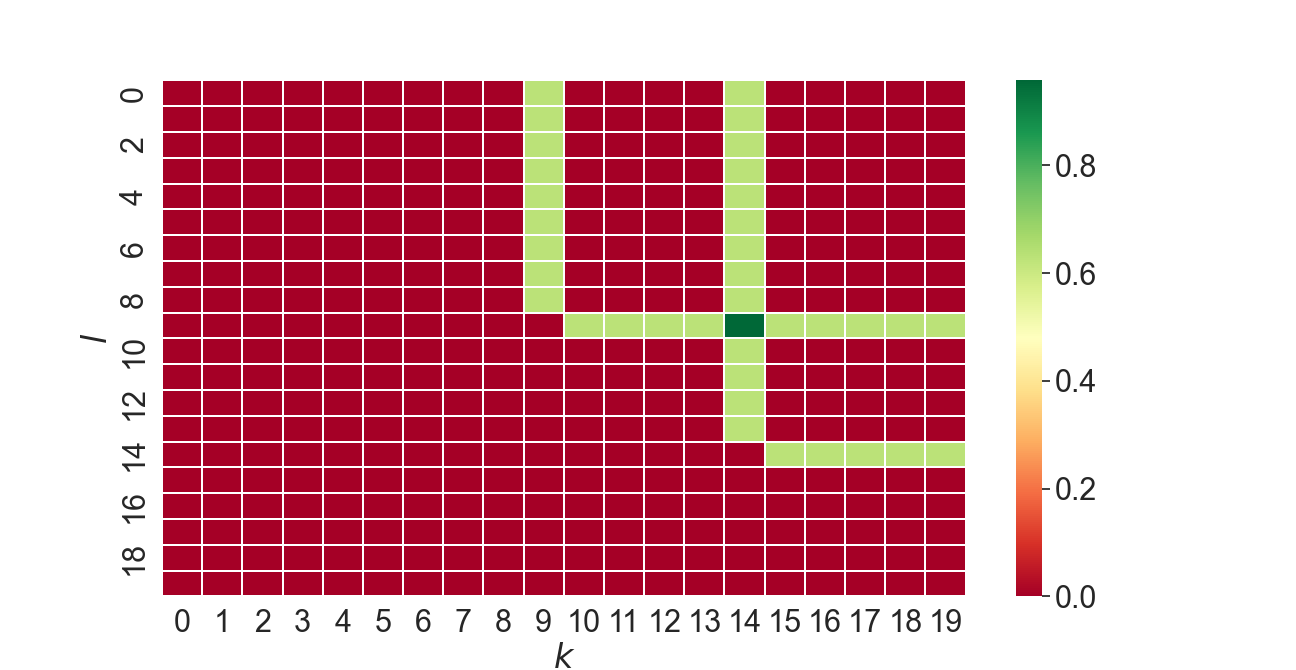}
    \caption{
    }
    \label{fig:second}
\end{subfigure}
\hfill
\begin{subfigure}{0.49\textwidth}
    \includegraphics[width=\textwidth]{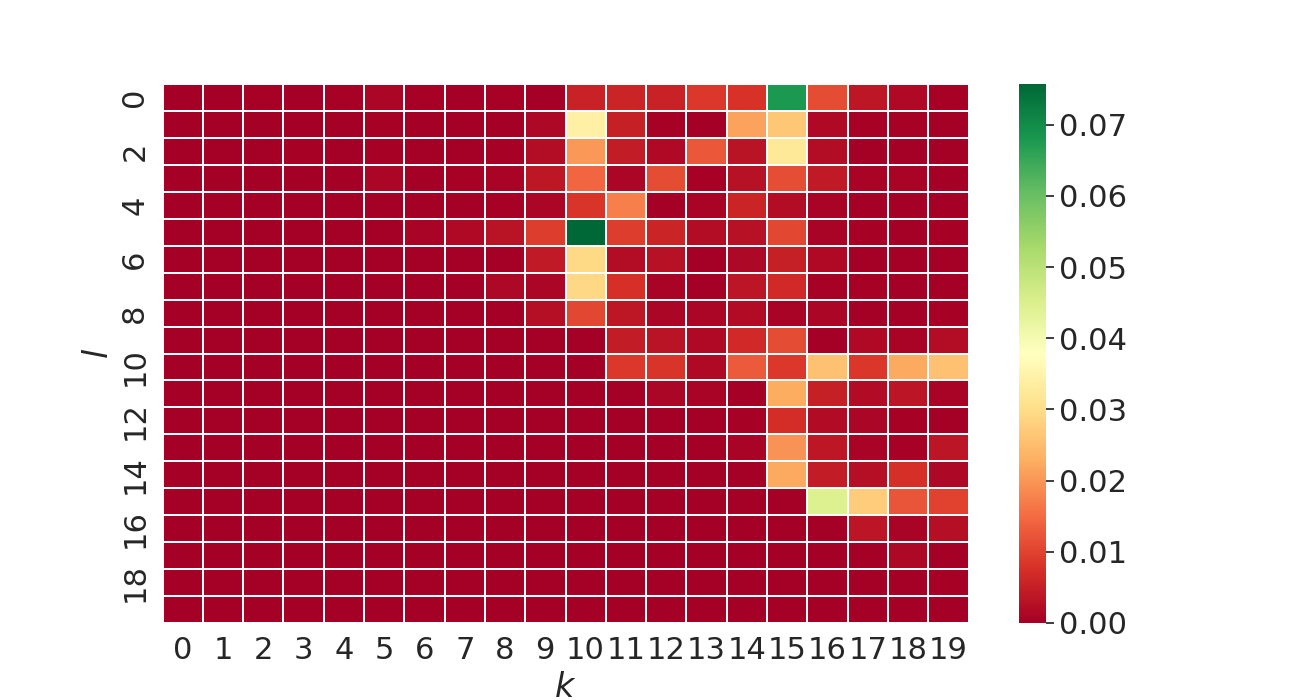}
 \caption{
    }
    \label{fig:fourth}
\end{subfigure}
\hfill\begin{subfigure}{0.49\textwidth}
    \includegraphics[width=\textwidth]{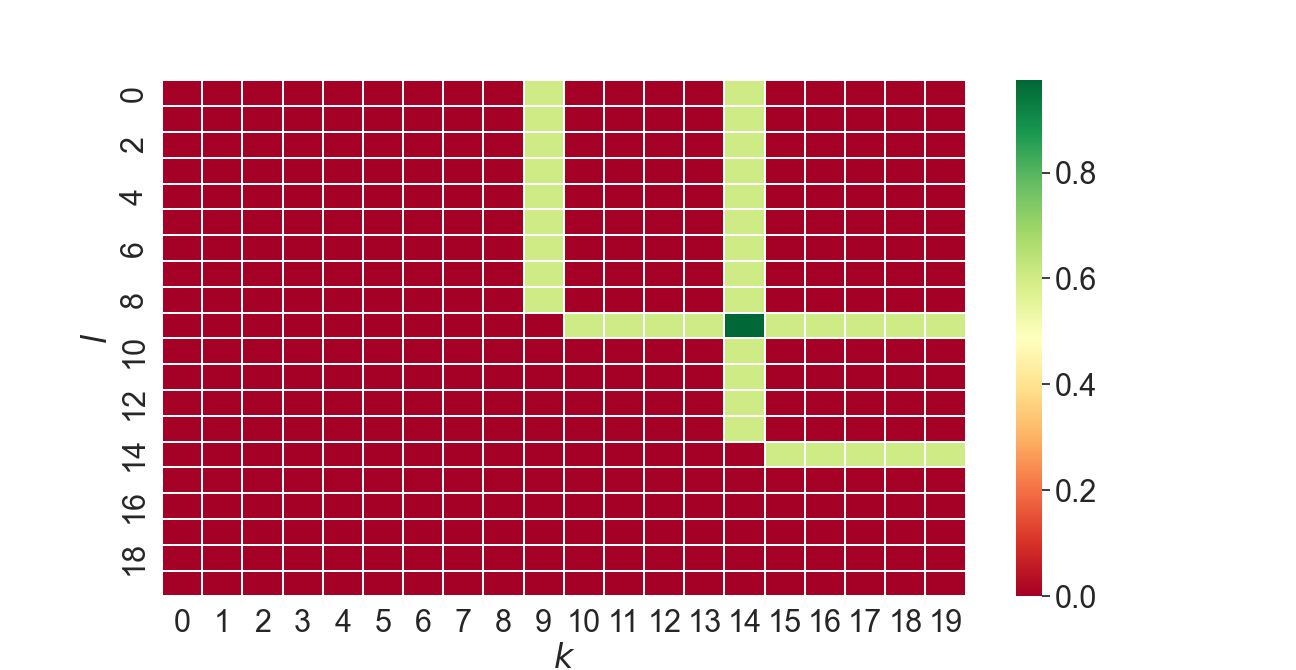}
    \caption{
    }
    \label{fig:second}
\end{subfigure}
\hfill
\begin{subfigure}{0.49\textwidth}
    \includegraphics[width=\textwidth]{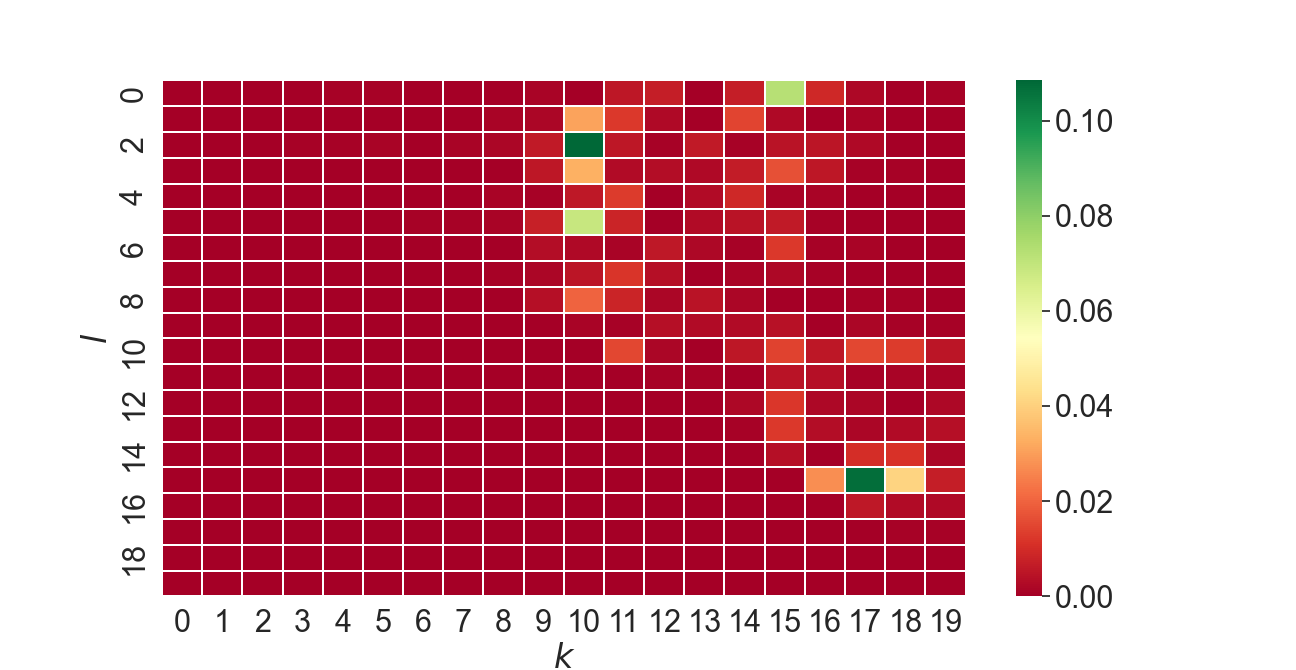}
    \caption{
    }
    \label{fig:fourth}
\end{subfigure}
\hfill
\begin{subfigure}{0.6\textwidth}
   \includegraphics[width=\textwidth]{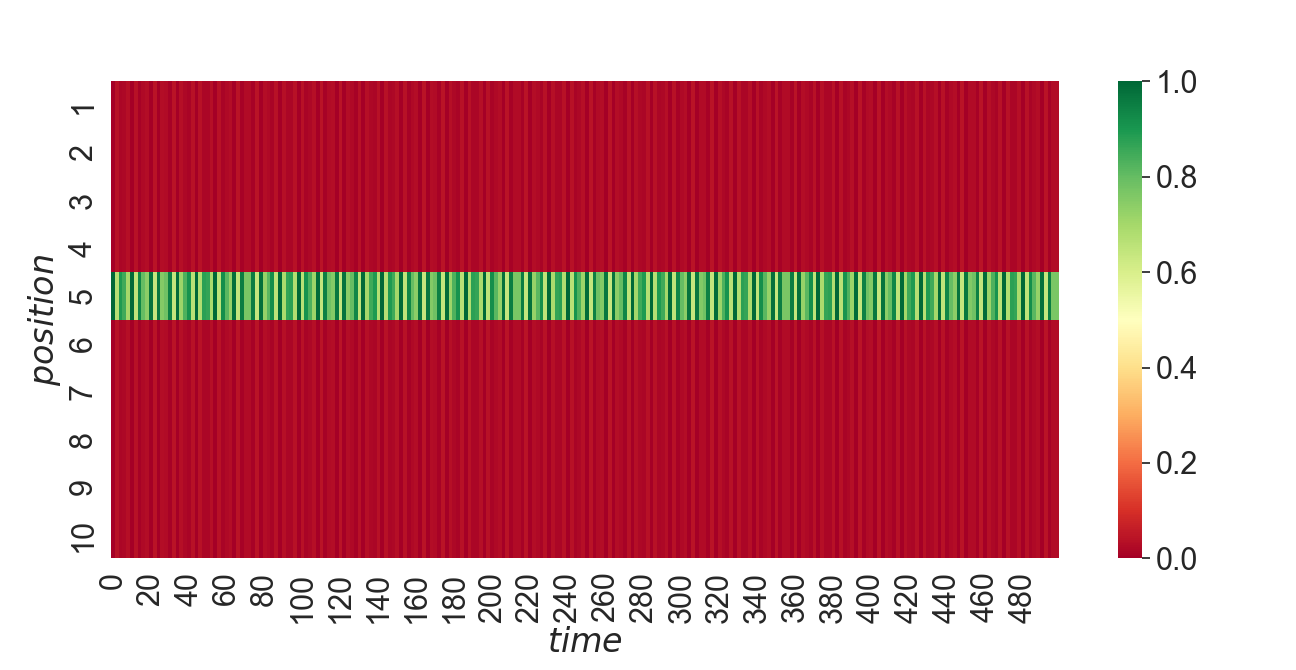}
    \caption{}
\label{fig:first}
\end{subfigure}
        
\caption{The probability distributions for the symmetric Hamiltonian, \eqref{eq:2-cycleH-alltoall} and the tight-binding Hamiltonian,  $\sum\limits_{j=1}^N~a^\dag_j a_{j+1} + h.c$. Localisation is observed for the former. (a) 2-fermion walk in the symmetric case. Probability for initial position ($l=9, k=14$) at $t = 3$. (b) 2-fermion walk in the tight-binding case. Probability for initial position ($l= 9, k=14$), at $t = 3$. (c) 2-fermion walk in the symmetric case. Probability for initial position ($l=9,k=14$), at $t = 3000$. (d) 2-fermion walk in the tight-binding case. Probability for initial position ($l=9,k=14$), at $t = 3000$. (e) 1-fermion walk in the symmetric case. Initial position = 5. }
\label{fig:probdistr}
\end{figure}

The non-zero probability distributions are found to be,
\begin{eqnarray}
    |\langle 1|1(t)\rangle|^2 & = & \frac{1}{N^2}\left[1+\left(N-1\right)^2 + 2\left(N-1\right)\cos{(Nt)}\right], \label{eq:1-fermionprobdistr1} \\
    |\langle j|1(t)\rangle|^2 & = & \frac{2}{N^2}\left[1-\cos{(Nt)} \right],\label{eq:1-fermionprobdistrj}
\end{eqnarray}
for $j\in\{2,3,\cdots, N\}$ and $\ket{j(t)} = e^{-\mathrm{i}Ht}\ket{j}$ are the time evolved states. The non-oscillating terms of both these expressions highlight the localisation effect. For large $N$ the term in \eqref{eq:1-fermionprobdistr1} goes to 1 whereas the term in \eqref{eq:1-fermionprobdistrj} goes to 0. These features are illustrated in Fig. \ref{fig:first}. 

The reason for the restricted evolution can also be seen from the explicit structure of the unitary evolution operator in the 1-fermion sector and this is shown in App. \ref{app:timeevolution1-fermionsector}. 

The phase of the oscillating term in \eqref{eq:1-fermionprobdistr1} and \eqref{eq:1-fermionprobdistrj} is the difference between the two energy levels in the 1-fermion sector. We have seen earlier that the addition of higher order interaction terms \eqref{eq:ftransposition-fullspace}, will only shift these energy levels of the Hamiltonian \eqref{eq:2-cycleH-alltoall} leaving the structure of the eigenstates intact. This implies that 
the localisation seen here is stable to the inclusion of such $\mathcal{S}_N$-symmetric interactions. This argument continues to hold even in the $k$-fermion sector as there are just two energy levels in each fermion number sector.

\paragraph{2-fermion walks :}
Using the eigenstates in \eqref{eq:Aalphaeigenstates} the amplitude for an initial 2-fermion state, $\ket{i,j}$ to end up in a state, $\ket{k,l}$ after a time $t$ is found to be,
\begin{eqnarray}\label{eq:2-fermionAmplitude}
    \vev{k,l|i,j(t)} & = & \frac{1}{N^2}\left[e^{-\mathrm{i}\left(N-2\right)t} \sum\limits_{\alpha=1}^{N-1}\left(\omega^{i\alpha}-\omega^{j\alpha} \right)\left(\omega^{-k\alpha}-\omega^{-l\alpha}\right) \right.\nonumber \\ 
    & + & \left. e^{2\mathrm{i}t}\mathop{\sum_{\alpha, \beta=1}^{N-1}}_ {\alpha<\beta}\left(\omega^{i\alpha + j\beta} - \omega^{i\beta + j\alpha} \right)\left(\omega^{-k\alpha - l\beta} - \omega^{-l\alpha - k\beta} \right)\right],
\end{eqnarray}
where $\ket{i,j(t)}$ are the time evolved 2-fermion states.
As mentioned earlier the restricted evolution is not apparent from \eqref{eq:2-fermionAmplitude}, but it becomes more transparent in a changed basis for the 2-fermion states\footnote{The orthogonality and completeness of these states is discussed in App. \ref{app:Completeness}.}. Consider the normalized eigenstates,
\begin{eqnarray}
    \ket{[j]}_{2} & = & \frac{1}{\sqrt{N-1}}~a_j^\dag \mathop{\sum_{i=1}^{N}}_ {i\neq j}~a_i^\dag\ket{\Omega}~;~j\in\{1,2,\cdots, N-1 \}, \label{eq:2-fermionji}\\
    \ket{[jkN]}_{2} & = & \frac{1}{\sqrt{3}}~\left(a_j^\dag a_k^\dag + a_k^\dag a_N^\dag + a_N^\dag a_j^\dag \right)\ket{\Omega}~;~j<k\in\{1,2,\cdots, N-1 \}.\label{eq:2-fermionjkN}
\end{eqnarray}
For these two sets of states, the notation $\ket{[~]}_{2}$ indicates that it is a linear combination of 2-fermion states.
From these expressions we see that there are $N-1$ eigenstates of this form in \eqref{eq:2-fermionji} and they come with the eigenvalue $(N-2)$ and there are $\frac{(N-1)(N-2)}{2}$ states of the form \eqref{eq:2-fermionjkN} with eigenvalue -2. This is consistent with the previous solution. As in the 1-fermion sector, the $\mathcal{S}_N$ symmetries, generated using \eqref{eq:ftransposition-2fermion}, map the degenerate eigenstates into each other. These operators can be written as products of the bilinears in $A_\alpha^\dag A_\beta$ and hence commute with the Hamiltonian in \eqref{eq:2-cycleHalpha}.

These eigenstates are used to expand $\ket{1,2}=a_1^\dag a_2^\dag\ket{\Omega}$ as
\begin{equation}\label{eq:2-fermion12}
    \ket{1,2} = \frac{\sqrt{N-1}}{N}\Big(\ket{[1]}_{2}-\ket{[2]}_{2}\Big) + \frac{\sqrt{3}(N-2)}{N}\ket{[12N]}_{2} - \frac{\sqrt{3}}{N}\sum\limits_{j=3}^{N-1}\Big(\ket{[1jN]}_{2} - \ket{[2jN]}_{2}\Big).
\end{equation}
The first three states in the above expression, $\ket{[1]}_{2}$, $\ket{[2]}_{2}$ and $\ket{[12N]}_{2}$, are linear combinations of $\ket{1,j}$, $\ket{2,j}$ with $j\in\{1,\cdots, N\}$. The states under the summation $\ket{[1jN]}_{2}$,  $\ket{[2jN]}_{2}$, also contains the $\ket{j,N}$ states but these cancel while taking the difference of these two states. Thus from these arguments it is clear that the time evolved $\ket{1,2}$ state will not overlap with a state $\ket{j,k}$ where neither of $j$ or $k$ is $(1,2)$. This is verified in the plot for the probability distribution shown in Fig. \ref{fig:second}. This is to be contrasted with a similar plot for a Hamiltonian that is not permutation invariant (see Fig. \ref{fig:fourth}), where the two fermions can now be found in states that do not follow the constraint for the symmetrised case \footnote{A similar result for quantum walks on Cayley graphs of the symmetric group is in \cite{Gerhardt2003ContinuousTimeQW}.}. Subsequently we can also compute the probability distributions
\begin{eqnarray}
    |\langle 1,2|1,2(t)\rangle|^2 & = & \frac{1}{N^2}\left[4+\left(N-2\right)^2 + 4\left(N-2\right)\cos{Nt}\right], \label{eq:2-fermionprobdistr12} \\
    |\langle \psi|1,2(t)\rangle|^2 & = & \frac{2}{N^2}\left[1-\cos{Nt} \right].\label{eq:2-fermionprobdistrPsi}
\end{eqnarray}
Here $\ket{\psi}$ denotes the allowed 2-fermion states that have an overlap with $\ket{1,2}$. These expressions present a clear indication of the localisation for large $N$ as the term $\frac{1}{N^2}\left[4+\left(N-2\right)^2\right]$ in \eqref{eq:2-fermionprobdistr12} tends to 1 and the term $\frac{2}{N^2}$ in \eqref{eq:2-fermionprobdistrPsi} approaches 0.  

\paragraph{$k$-fermion walks :}
This constraining feature continues to hold true for the amplitudes and the corresponding probability distributions in a general $k$-fermion sector. We will see below that a point in $k$-dimensional space occupied by $k$ fermions moves to points where at least $k-1$ of the coordinates coincide with the initial state. As in the 2-fermion case this becomes apparent when we work with $\frac{(N-1)\,!}{(k-1)\,!(N-k)\,!}$ eigenstates of the form, 
\begin{equation}\label{eq:k-fermionstates1set}
   \ket{\left[i_1i_2\cdots i_{k-1}\right]}_{k}=\frac{1}{\sqrt{N-k+1}}a_{i_1}^\dag a_{i_2}^\dag\cdots a_{i_{k-1}}^\dag \mathop{\sum_{j=1}^{N}}_ {j\neq \{i_1,i_2,\cdots i_{k-1}\}}a_j^\dag\ket{\Omega}, 
\end{equation}
with $i_1<i_2<\cdots <i_{k-1}\in\{1,\cdots N-1\}$.
The second set of eigenstates\footnote{The proof for this is in App. \ref{app:k-fermionstates2set}.} account for $\frac{(N-1)\,!}{k\,!(N-k-1)\,!}$ of them and take the form,
\begin{equation}\label{eq:k-fermionstates2set}
    \ket{\left[i_1i_2\cdots i_kN\right]}_{k}= \frac{1}{\sqrt{k+1}}\left[a_{i_1}^\dag\cdots a_{i_k}^\dag +\left(-1\right)^k a_{i_2}^\dag\cdots a_{N}^\dag + \cdots + \left(-1\right)^{k^2}a_N^\dag\cdots a_{i_{k-1}}^\dag\right]\ket{\Omega}.
\end{equation}
An initial state of the form $\ket{1,2,\cdots, k}$ can be expanded using the above eigenstates such that each of them contain at least $k-1$ of $\{ 1,2,\cdots k\}$. Finally the probability distribution for the time evolved $k$-fermion state to overlap with its initial state is,
\begin{equation}
    |\langle 1,2,3,\cdots, k|1,2,3,\cdots, k(t)\rangle|^2 = \frac{1}{N^2}\left[ k^2 + \left(N-k\right)^2 + 2k\left(N-k\right)\cos{Nt}\right],
\end{equation}
generalising the result in \eqref{eq:2-fermionprobdistr12}. It is clear from this expression that the localisation feature continues to hold for the $k$-fermion states as well. 

\section{Discussion}
\label{sec:discussion}
We have explored the question of localisation in a system with superselection sectors and a global discrete symmetry. We saw that in a fermionic system with a global $\mathcal{S}_N$ symmetry all the basis states of the many-fermion Hilbert space completely localise at all times, $t$ for large $N$.

We can understand this result in more general terms as follows. Consider a finite dimensional quantum system parametrised by the tuples, $\lambda_j=\{N, \theta, k, \cdots \}$ with the dimension of the Hilbert space being a function of the integer $N$. The $\theta$'s can be taken as complex in general and the $k$'s could denote another set of integers. The $j$'s index the eigenvalues which, for a system with huge degeneracies, is expected to be much less than the size of the Hilbert space. Furthermore there could be even more parameters denoted by the $\cdots$, but for the examples we will consider here these three sets will suffice. We expect such parameters to appear in the Hamiltonian describing the evolution of such quantum systems. 

For example, the fully symmetric Hamiltonian \eqref{eq:2-cycleH-alltoall} considered in this paper is only parametrised by $N$ and that presents the only scale for this system. Other systems with lesser and no symmetries will include more parameters. In all such systems the probability distributions are computed using the overlaps
\begin{equation}\label{eq:clambdaj}
    \vev{\psi|\psi(t)} = \sum\limits_j~|c_{\lambda_j}|^2e^{-\mathrm{i}E_{\lambda_j}t},
\end{equation}
where the $E_{\lambda_j}$'s are the energy eigenvalues parametrised by the $\lambda_j$'s. The normalisation factors in the system are expected to go as $O(\frac{1}{N})$ and this is carried over to the coefficients, $|c_{\lambda_j}|^2$. The structure of the coefficients will determine the scaling of these overlaps for large $N$. In the case where $N$ appears as the lone parameter in the system, the degeneracies of the eigenvalues will determine the localisation properties. We do not expect to see localisation if the number of eigenvalues are of $O(N)$ or the degeneracies are of $O(1)$. On the other hand when there are very few eigenvalues with some of them having degeneracies of $O(N)$ we will certainly see localisation. 

This is precisely the scenario for the $\mathcal{S}_N$-symmetric system studied earlier. When the system includes more parameters, the coefficients $c_{\lambda_j}$ will also depend on them. In such cases these parameters can be modulated to determine the localisation properties for fixed values of $N$. The energy eigenvalues only appear in the phases of the oscillating terms and are thus not affected at large $N$.

We will explore such scenarios in this section which will show that this type of localisation can occur in more general systems than the highly symmetric case studied earlier. Furthermore we also propose a prescription for this type of disorder-free localisation. To this end we will elaborate on the following points that will help construct other systems that show a similar phenomenon of disorder-free localisation.
\begin{enumerate}
    \item Including perturbations that break the global $\mathcal{S}_N$ symmetry. We will discuss the cases when localisation persists and when it is destroyed.
    \item Sensitivity of the system to initial conditions. Do the superposition of basis elements of each number sector also show localisation? This is another possibility that can affect the coefficients, $|c_{\lambda_j}|$. Here again we will see under what conditions localisation survives and which states delocalise.
    \item Role played by the symmetries in the localisation. Is this localisation only true for fermionic systems? Do spin chains and bosonic systems with global $\mathcal{S}_N$ symmetry also exhibit this type of disorder-free localisation? Do the localisation features continue to hold for models with reduced global symmetry, namely $\mathcal{S}_{N-k}$?
    \item We suggest a plausible prescription by noting a connection between graph theory and the disorder-free localisation studied here. 
\end{enumerate} 

\subsection{Effect of perturbations}
\label{subsec:perturbations}
The fully symmetric model features localisation at large $N$ and for all times $t$. Naively we expect a generic perturbation, that breaks the global $\mathcal{S}_N$ symmetry, to make the system delocalise at large times $t$. We now study the effects of symmetry breaking perturbations for our system by considering the inclusion of two types of terms - terms which are local and global with respect to the conjugate variables. More precisely by local we mean that the number of additional terms is much less than the $O(N)$ and by global we mean that the number of terms we include spans the full range of the conjugate variable, i.e. $N$. In the global case we find that the system can both localise and delocalise depending on the coefficients appearing in the perturbation while in the local case the system stays localised always. 
\paragraph{Case 1 :} We assume perturbations of the form $A_\alpha^\dag A_\beta$, $A_{\alpha_1}^\dag A_{\alpha_2}^\dag A_{\beta_1}A_{\beta_2}$, and higher order terms. For specific values of the conjugate indices $\alpha$'s and $\beta$'s these terms are clearly non-local in the real space indices $i$, $j$, $k$ etc, and they break the global $\mathcal{S}_N$ symmetry. For the first case we keep the number of such terms to be less than $O(N)$. They commute with the original Hamiltonian and hence they introduce only a small split in the energy levels. In particular the energy levels of say, the $k$-particle sector, will split into two groups centered around the eigenvalues $-k$ and $N-k$, and with further splits in each group. As a simple example consider the Hamiltonian,
\begin{equation}\label{eq:Htheta}
    H_\theta = N A_N^\dag A_N - \hat{N} + \theta~A_\alpha^\dag A_\beta + \theta^*~A_\beta^\dag A_\alpha;~~\alpha\neq\beta\neq N.
\end{equation}
The spectrum splits around the unperturbed energy eigenvalues and this is shown for the 1-fermion and $k$-fermion sectors in Tables \ref{tab:1-particletheta}, \ref{tab:k-particletheta}. In these cases the localisation of the basis states for all times continues to hold for any value of the interaction coefficients.
\begin{table}[h!]
\centering
\begin{tabular}{ |c|c|c| } 
 \hline
 Eigenvalue & Eigenstate & Degeneracy \\
 \hline
 $N-1$ & $A_N^\dag\ket{\Omega}$ & 1 \\ 
 \hline
 -1 & $A_\gamma^\dag\ket{\Omega};~\gamma\neq \alpha, \beta, N$ & $N-3$ \\ 
 \hline
 $|\theta|-1$ & $\frac{1}{\sqrt{2|\theta|}}\left[\sqrt{\theta}A_\alpha^\dag + \sqrt{\theta^*}A_\beta^\dag \right]\ket{\Omega}$ & 1 \\
 \hline
 $-|\theta|-1$ & $\frac{1}{\sqrt{2|\theta|}}\left[-\sqrt{\theta}A_\alpha^\dag + \sqrt{\theta^*}A_\beta^\dag \right]\ket{\Omega}$ & 1 \\
 \hline
 \end{tabular}
 \caption{The 1-particle spectrum of \eqref{eq:Htheta}.}
 \label{tab:1-particletheta}
\end{table}

\begin{table}[h!]
\centering
\begin{tabular}{ |c|c|c| } 
 \hline
 Eigenvalue & Eigenstate & Degeneracy \\
 \hline
 \hline
 $N-k$ & $A_N^\dag A_{\gamma_1}^\dag\cdots A_{\gamma_{k-1}}^\dag\ket{\Omega}$ & $\binom{N-3}{k-1}$ \\ 
 $N-k$ & $A_N^\dag A_\alpha^\dag A_\beta^\dag A_{\gamma_1}^\dag\cdots A_{\gamma_{k-3}}^\dag\ket{\Omega}$ & $\binom{N-3}{k-3}$\\
\hline
$N-k+|\theta|$ & $\frac{1}{\sqrt{2|\theta|}}A_N^\dag\left[\sqrt{\theta}A_\alpha^\dag + \sqrt{\theta^*}A_\beta^\dag \right]A_{\gamma_1}^\dag\cdots A_{\gamma_{k-2}}^\dag\ket{\Omega}$ & $\binom{N-3}{k-2}$\\
\hline
 $N-k-|\theta|$ & $\frac{1}{\sqrt{2|\theta|}}A_N^\dag\left[-\sqrt{\theta}A_\alpha^\dag + \sqrt{\theta^*}A_\beta^\dag \right]A_{\gamma_1}^\dag\cdots A_{\gamma_{k-2}}^\dag\ket{\Omega}$ & $\binom{N-3}{k-2}$\\
 \hline
 \hline
 $-k$ & $A_{\gamma_1}^\dag\cdots A_{\gamma_{k}}^\dag\ket{\Omega}$ & $\binom{N-3}{k}$ \\ 
 $-k$ & $A_\alpha^\dag A_\beta^\dag A_{\gamma_1}^\dag\cdots A_{\gamma_{k-2}}^\dag\ket{\Omega}$ & $\binom{N-3}{k-2}$ \\ 
 \hline
 $|\theta|-k$ & $\frac{1}{\sqrt{2|\theta|}}\left[\sqrt{\theta}A_\alpha^\dag + \sqrt{\theta^*}A_\beta^\dag \right]A_{\gamma_1}^\dag\cdots A_{\gamma_{k-1}}^\dag\ket{\Omega}$ & $\binom{N-3}{k-1}$ \\
 \hline
 $-|\theta|-k$ & $\frac{1}{\sqrt{2|\theta|}}\left[-\sqrt{\theta}A_\alpha^\dag + \sqrt{\theta^*}A_\beta^\dag \right]A_{\gamma_1}^\dag\cdots A_{\gamma_{k-1}}^\dag\ket{\Omega}$ & $\binom{N-3}{k-1}$ \\
 \hline
 \end{tabular}
 \caption{The $k$-particle spectrum of \eqref{eq:Htheta}.}
 \label{tab:k-particletheta}
\end{table}

The probability amplitude for the time 0 state, $|i\rangle:=a^{\dagger}_i|\Omega\rangle$ to be still in $|i\rangle$ after a time $t$ is found to be
\begin{eqnarray}
\begin{split}
    \langle i | i(t)\rangle &=& \frac{1}{N}\Big[(N-3)e^{it}+e^{-i(N-1)t}+2 \cos^2\left(\frac{\pi}{N}(\beta-\alpha)+\frac{\phi}{2}\right)e^{-i(|\theta|-1)t} +\\  && 2\sin^2\left(\frac{\pi}{N}(\beta-\alpha)+\frac{\phi}{2}\right)e^{i(|\theta|+1)t}\Big]
\end{split}
\end{eqnarray}
with $\phi$ being the phase factor of the complex number $\theta$. An analogous computation for the $k$-particle sector of \eqref{eq:Htheta} gives,
\begin{equation}\label{eq:ratiotheta}
  \langle i_1\cdots i_k|i_1\cdots i_k(t)\rangle = \frac{1}{\binom{N}{k}}\left[\binom{N-3}{k} + \binom{N-3}{k-2}\right]e^{\mathrm{i}kt} + O\left(\frac{1}{N}\right).  
\end{equation}
It is easily verified that the binomial coefficient in \eqref{eq:ratiotheta} goes to unity in the large $N$ limit and this is precisely the signature of localisation as seen in the fully $\mathcal{S}_N$ symmetric case. 
\paragraph{Quartic interaction:} Up to now, we only have considered perturbations which are quadratic in the creation, annihilation operators. Now let us introduce a quartic interaction which modifies the Hamiltonian to
\begin{eqnarray}\label{eq:quartic H}
    H_{\theta}=NA_N^{\dagger}A_N-\hat{N}+\theta A^{\dagger}_{\alpha}A^{\dagger}_{\beta}A_{\beta}A_{\alpha};\quad\quad \alpha\neq\beta\neq N
\end{eqnarray}
As before, this interaction which is seemingly local in the conjugate variables, $\alpha$'s, become global in the original indices labelled by the $j$'s and also breaks the global $\mathcal{S}_N$ symmetry. Addition of this term affects every $k\geq 2$ particle sector. The spectral information in an arbitrary sector can easily be extracted and is tabled below.

\begin{table}[h!]
\centering
\begin{tabular}{ |c|c|c| } 
 \hline
 Eigenvalue & Eigenstate & Degeneracy \\
 \hline
 \hline
 $N-k$ & $A_N^\dag A_{\gamma_1}^\dag\cdots A_{\gamma_{k-1}}^\dag\ket{\Omega}$ & $\binom{N-3}{k-1}$ \\ 
 $N-k$ & $A_N^\dag A_\alpha^\dag A_{\gamma_1}^\dag\cdots A_{\gamma_{k-2}}^\dag\ket{\Omega}$ & $\binom{N-3}{k-2}$\\
 $N-k$ & $A_N^\dag A_\beta^\dag A_{\gamma_1}^\dag\cdots A_{\gamma_{k-2}}^\dag\ket{\Omega}$ & $\binom{N-3}{k-2}$\\
\hline
$N-k+\theta$ & $A_N^\dag A_\alpha^\dag A_\beta^\dag A_{\gamma_1}^\dag\cdots A_{\gamma_{k-3}}^\dag\ket{\Omega}$ & $\binom{N-3}{k-3}$\\
\hline
 \hline
 $-k$ & $A_{\gamma_1}^\dag\cdots A_{\gamma_{k}}^\dag\ket{\Omega}$ & $\binom{N-3}{k}$ \\ 
 $-k$ & $A_\alpha^\dag A_{\gamma_1}^\dag\cdots A_{\gamma_{k-1}}^\dag\ket{\Omega}$ & $\binom{N-3}{k-1}$ \\ 
  $-k$ & $A_\beta^\dag A_{\gamma_1}^\dag\cdots A_{\gamma_{k-1}}^\dag\ket{\Omega}$ & $\binom{N-3}{k-1}$ \\ 
 \hline
 $\theta-k$ & $\ A_\alpha^\dag A_\beta^\dag A_{\gamma_1}^\dag\cdots A_{\gamma_{k-2}}^\dag\ket{\Omega}$ & $\binom{N-3}{k-2}$\\
 \hline
 \end{tabular}
 \caption{The $k$-particle spectrum of \eqref{eq:quartic H}.}
 \label{tab:k-particletheta}
\end{table}
The probability amplitude for the $k$-particle sector of \eqref{eq:quartic H} gives,
\begin{equation}\label{eq:ratiotheta}
  \langle i_1\cdots i_k|i_1\cdots i_k(t)\rangle = \frac{1}{\binom{N}{k}}\left[\binom{N-3}{k} + 2\binom{N-3}{k-1}\right]e^{\mathrm{i}kt} + O\left(\frac{1}{N}\right).  
\end{equation}
In the large $N$ limit, the modulus goes to unity and therefore the state localizes. Thus these perturbations that break the global $\mathcal{S}_N$ symmetry do not spoil the localisation features of the fully symmetric system. This continues to hold as long the number of such perturbations is not too many or is much less than the $O(N)$ and breaks down when the number of terms is comparable to $N$. This brings us to the second case.

\paragraph{Case 2 :} For the second case we continue including similar interactions as in the first case but now the number of such terms included are of $O(N)$. For example, $\sum\limits_{\alpha=1}^N~\epsilon_\alpha A_\alpha^\dag A_\alpha $ is one such simple term. We see that such terms do not disturb the eigenstates and the energy levels continue to split into two groups as long as the perturbation coefficients $\epsilon_\alpha$, are not comparable to $N$. Assuming this is the case the split is much finer compared to the previous situation and hence the degeneracy drastically reduces and is no longer of the $O(N)$ when the $\epsilon_\alpha$'s are all unequal. However if a significant number of $\epsilon_\alpha$'s are equal to each other we expect to see a different behavior. We will now illustrate both these situations.

\begin{figure}[h]
    \centering
    \includegraphics[scale=0.9]{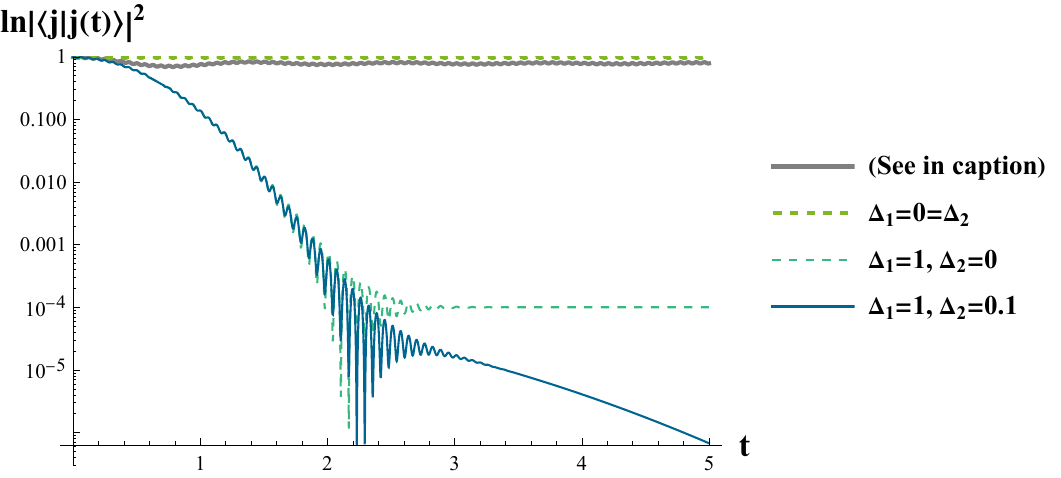}
    \caption{(Color Online) The plot of the probability distribution $\ln{|\langle j|j(t)\rangle|^2}$ in the $1$-particle sector for $N=100$ with energy splitting as given in Case 2. The thick, gray curve represents the case where $90$ out of the $100$ eigenvalues are shifted by an equal amount of $1$ and the rest $10$ eigenvalues are shifted by amounts $2\sqrt{l},l=(1,2,\cdots,10)$. The other three curves follow from the expression (\ref{1-Gaussian}) with specific values of $\Delta_1$ and $\Delta_2$ as prescribed in the figure. As discussed below, this also can be thought of representing the behaviour of the \textit{Spectral Form Factor}.}
    \label{fig:SFF1}
\end{figure}

We consider the case $G\gg\Delta$, where $\Delta$ is a measure of the energy splitting caused by the perturbation and $G$ is the degeneracy before the splitting. Further we demand that no two energy levels occupy same energy. Now owing to the presence of a large number of eigenstates within a very small interval of energy, it is possible to have continuous energy levels and thus an energy distribution function can be associated with them. Also the distribution function $\rho(E)$ should be such that $\rho(E)|_{\Delta\to 0}$ gives the correct result for the unsplit case. One possible choice can be
\begin{eqnarray}
    \rho(E):= \frac{G}{\Delta\sqrt{\pi}}\exp\left(-\frac{(E-E_0)^2}{\Delta^2}\right)
\end{eqnarray}
with $E_0$ being the energy before splitting. The corresponding amplitude in the one-particle sector becomes
\begin{eqnarray}\label{1-Gaussian}
    \langle i|i(t)\rangle=\frac{1}{N^2}\left[(N-1)e^{it-\frac{\Delta_1^2t^2}{4}}+e^{-i(N-1+\Delta_2)t}\right]
\end{eqnarray}
with $\Delta_1$ and $\Delta_2$ being the shifts in the energy levels $-1$ and $N-1$ respectively. In general for $k$-particle sector, we have the expression
\begin{eqnarray}\label{k-Gaussian}
    \langle i_1\cdots i_k|i_1\cdots i_k(t)\rangle=\frac{1}{\binom{N}{k}}\left[\binom{N-1}{k}e^{\text{i}kt-\frac{\Delta_1^2t^2}{4}}+\binom{N-1}{k-1}e^{-\text{i}(N-k)t-\frac{\Delta_2^2t^2}{4}}\right]
\end{eqnarray}
where $\Delta_1$ and $\Delta_2$ are the spreads in the energies $k$ and $N-k$ respectively. We have shown the result graphically in Fig.\ref{fig:SFF1}.

In contrast to this, let us consider another situation  where we essentially continue working with the previous interaction term, but now $m$ number of $\epsilon_{\alpha}$'s assume exactly the same value, with $m\sim O\left(N\right)$. As argued earlier, this suggests localization and is further verified by the thick gray plot in Fig.\ref{fig:SFF1}.

Interestingly enough, one arrives at identical results while exploring the so-called \textit{spectral form factor} (SFF), whose behaviour is believed to indicate thermalisation and chaos in quantum many-body systems \cite{PhysRevE.55.4067,cotler2017black,cipolloni2023spectral}. The SFF, denoted by $\mathcal{F}(t)$, is defined as 
\begin{eqnarray}\label{SFF}
    \mathcal{F}(t)=\Big|\frac{\mathcal{Z}(it)}{\mathcal{Z}(0)}\Big|^2
\end{eqnarray}
where $\mathcal{Z}(\beta+it)=\text{Tr}\left[e^{-\beta H - i H t}\right]$ is the analytically continued partition function. At $\beta=0$, the ratio of the partition functions in the $k$-particle sector becomes 
\begin{eqnarray}
\frac{\mathcal{Z}(it)}{\mathcal{Z}(0)}=\frac{\text{Tr}\left[e^{-i H t}\right]}{\text{Tr}[I]}=\frac{1}{\binom{N}{k}}\sum_{\alpha} e^{-i E_{\alpha} t}
\end{eqnarray}
which, once appropriate energy distributions are chosen, turns out to be exactly the same expression given in (\ref{k-Gaussian}). This happens precisely because the states $|i_1\cdots i_k\rangle$'s are equal superpositions of the eigenstates of the Hamiltonian. Therefore the analysis of the SFF for this particular case can capture the existing features equally well.

Note that we have used number preserving perturbations and those that commute with the unperturbed Hamiltonian. Perturbations that do not possess these two properties require further investigation.

\subsection{Sensitivity of the system to initial conditions}
\label{subsec:initialconditions}
Using the time evolution operator in the one-particle sector (See \ref{app:timeevolution1-fermionsector}) we investigate the sensitivity of the system to the initial conditions by considering an arbitrary initial state $|\psi\rangle = \sum_{j}c_j |j\rangle$, with $\sum\limits_{j}|c_j|^2=1$. 
The quantity $|\langle\psi|\psi(t)\rangle|^2$ turns out to be
\begin{eqnarray}\label{prob_arb_iniit-state}
   \mathcal{P}(t):=|\langle\psi|\psi(t)\rangle|^2 &=& 1-\frac{2 \tau (N-\tau)}{N^2}\left[1-\cos{(Nt)}\right]
\end{eqnarray}
with  $\tau:=\sum_{i,j}\Bar{c_i}c_j=\bar{C}C; \;C:=\sum_j c_j$. Evidently the quantity $\tau$ is strictly positive and we can find the maximum value it assumes by the method of Lagrange multipliers. This amounts to solving the system of equations given by $\frac{\partial \tau}{\partial c_i}=\lambda\frac{\partial g}{\partial c_i}$; with $ g=\sum_i |c_i|^2-1$ and $\lambda$ being the Lagrange multiplier. The maximum value can be found as $\tau_{max}=N$. 

\begin{figure}[h!tbp]
    \centering
    \includegraphics[scale=0.7]{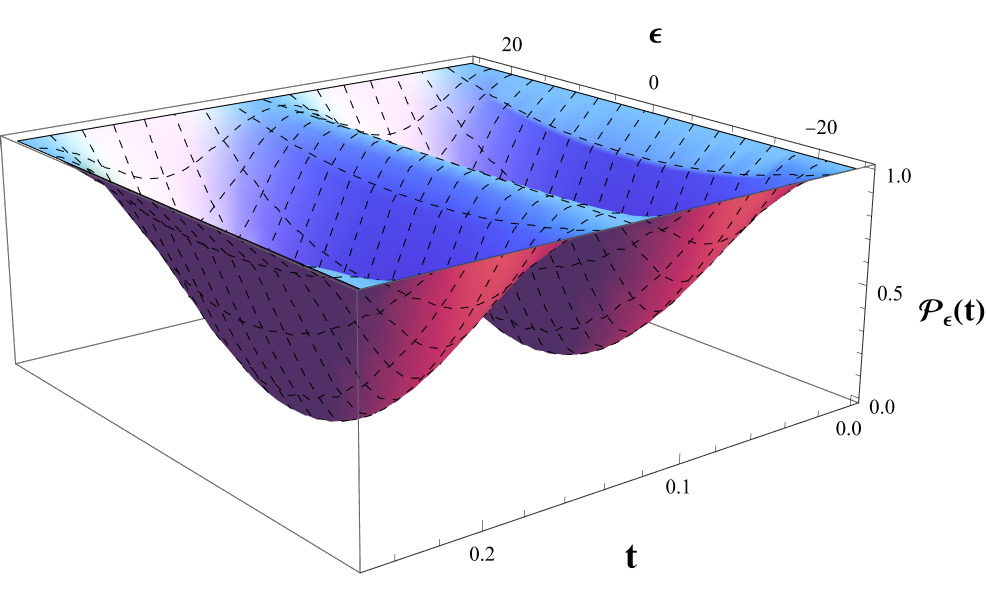}
    \caption{$\mathcal{P}_{\epsilon}(t)$ as function of $\epsilon$ and $t$ ($N=50$)}
    \label{fig:sensitivty}
\end{figure}

Note that $\mathcal{P}(t)$ is symmetric about $\tau=N/2$. Around $\tau=\frac{N}{2}+\epsilon$ this becomes
\begin{eqnarray}
    \mathcal{P}_{\epsilon}(t)=\Big(\frac{1}{2}+\frac{2 \epsilon^2}{N^2}\Big)+\Big(\frac{1}{2}-\frac{2 \epsilon^2}{N^2}\Big)\cos{(N t)}
\end{eqnarray}
When $\epsilon=0$, we have $ \mathcal{P}_0(t)=\frac{1}{2}(1+\cos{(Nt)})$ and this clearly is not localized. On the other hand, if $\epsilon=\pm\frac{N}{2}$, we find $\mathcal{P}_{\pm\frac{N}{2}}(t)=1$ and the state is completely localized. In fact, $\tau =N $ corresponds to the eigenstate $|\Phi\rangle=\frac{1}{\sqrt{N}}\sum_i a^{\dagger}_i|\Omega \rangle$ and $\tau=0$ corresponds to the eigenstates of the form $|\phi_{[i,j]}\rangle =\frac{1}{\sqrt{2}}\left(a_i^{\dagger}-a_j^{\dagger}\right)|\Omega\rangle$. If we plot $\mathcal{P}_{\epsilon}(t)$ as a function of both $\epsilon$ and $t$, we obtain the plot in Fig. \ref{fig:sensitivty}.
One can see for $\epsilon\sim\pm\frac{N}{2}$ the probability remains very close to unity, i.e. the corresponding states are localized. Whereas near $\epsilon = 0$ the probability oscillates with time from zero to unity, suggesting that those states are not localized at all. 

The result given by (\ref{prob_arb_iniit-state}) is also valid for the case of generic $k$-particle sector. The quantity $\tau$ turns out to be some complicated combination of the coefficients $c$'s, appearing in the expansion of the state $|\Psi\rangle$ in the basis $|i_1\cdots i_k\rangle$. The phenomenon of localization continues to depend on the quantity $\tau$ in the same way as in the case of one-particle sector.

\subsection{Role of symmetries in the localisation}
\label{subsec:roleofsymmetries}
It is important to understand the source of the localisation and the role played by the two symmetries, the superselected symmetry and the global $\mathcal{S}_N$ symmetry, in obtaining this feature. To this end we consider the following possibilities. 
\begin{enumerate}
    \item Systems with reduced global symmetry, 
$\mathcal{S}_{N-k}$. We analyse the $k=1$ case in full detail and then consider the general $k$ case. We find that all these systems localise in a manner similar to the fully symmetric case implying that the global $\mathcal{S}_N$ symmetry, though sufficient, is not necessary to obtain these localisation features.
     \item  Next we check if this feature is exclusive to fermionic systems and to verify this we find similar localisation features in a spin chain system with global $\mathcal{S}_N$ symmetry and also in $\mathcal{S}_N$ symmetric bosonic systems.
\end{enumerate}

\begin{figure}[h!]
\centering
\begin{subfigure}{.5\textwidth}
  \centering
  \includegraphics[width=1\linewidth]{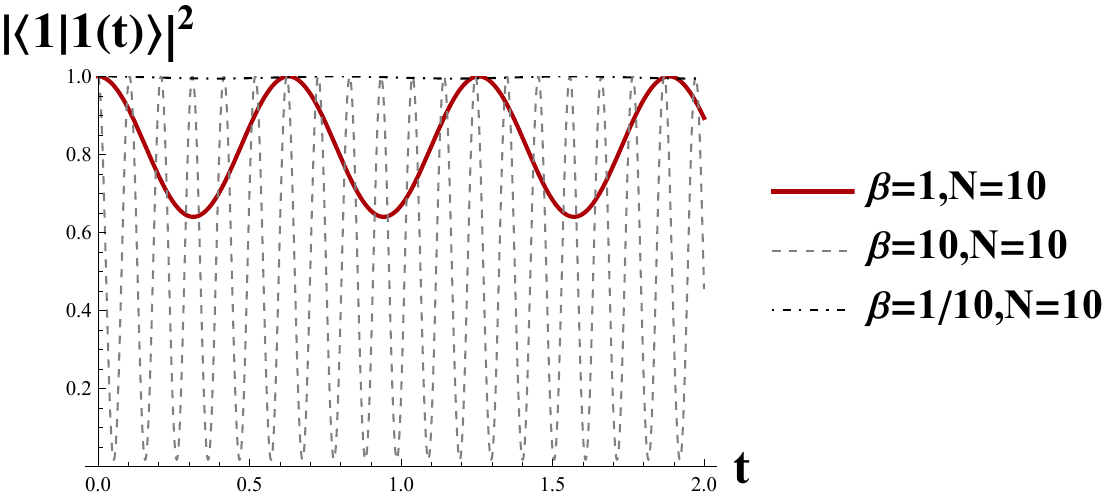}
  \caption{}
  \label{fig:p1}
\end{subfigure}%
\begin{subfigure}{.5\textwidth}
  \centering
  \includegraphics[width=1\linewidth]{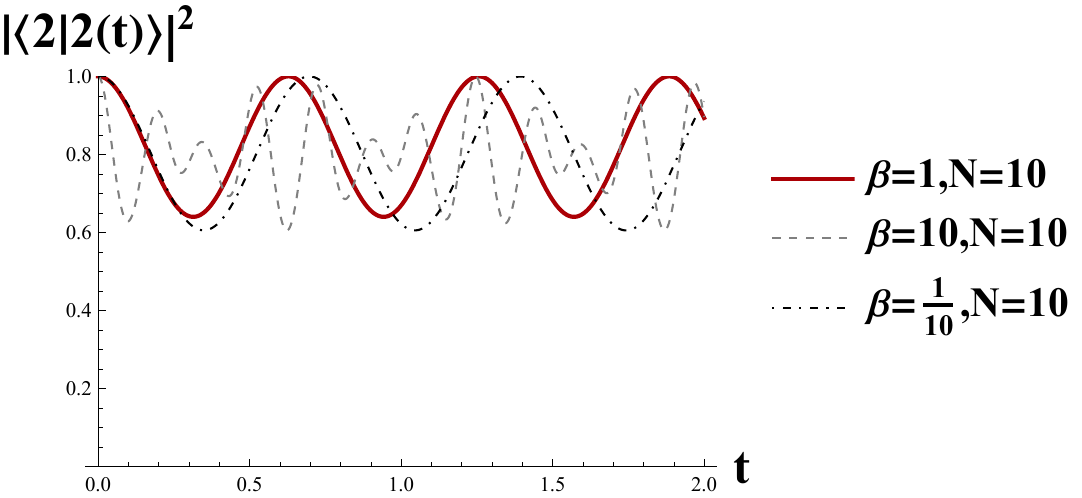}
  \caption{}
  \label{fig:p2}
\end{subfigure}
\caption{(Color Online) Probability distributions for the 1-fermion sector of \eqref{eq:markedHamiltonian} where the 1st site is marked. (a) $|\langle 1|1(t)\rangle|^2$ vs $t$, and (b) $|\langle 2|2(t)\rangle|^2$ vs $t$. The $\beta=1$ curves correspond to the fully symmetric case and is the same in both (a) and (b). The first particle can now be localised for even small values for $N$ by tuning $\beta$ appropriately ($\beta=\frac{1}{10}$ curve of (a)). The other sites with the $\mathcal{S}_{N-1}$ symmetry will localise for large $N$. When $\beta$ is comparable to the value of $N$ ($\beta=N$ curves) we do not see localisation.}
\end{figure}

\paragraph{Models with reduced global symmetry - $\mathcal{S}_{N-1}$ :} In the first situation we continue working with fermionic systems and reduce the explicit global $\mathcal{S}_N$ symmetry. This is done by `marking' a single site, say 1, to modify the Hamiltonian in \eqref{eq:2-cycleH-alltoall} to,
\begin{equation}\label{eq:markedHamiltonian}
   H = \beta\sum\limits_{j=2}^N~\left[a_1^\dag a_j + a_j^\dag a_1 \right] + \mathop{\sum\limits_{j,k=2}^N}_{j<k}~\left[a_k^\dag a_j + a_j^\dag a_k \right].
\end{equation}
This model has a global $\mathcal{S}_{N-1}$ symmetry among the sites $\{2,3,\cdots, N \}$\footnote{These models can be related to {\it central spin systems} \cite{Arenz2013ControlOO, Nepomechie2018TheSH, Villazon2020PersistentDS} when the non-marked sites are not interacting with each other. These Hamiltonians are also related to the adjacency matrices of {\it cone graphs} where similar localisation features are studied \cite{Godsil2017SedentaryQW, Carlson2006UniversalMO} (See Sec. \ref{subsec:DFLgraphs} for the connection to graph theory).}. To analyse the consequences we rewrite the Hamiltonian in a different fashion. In the 1-fermion sector spanned by 
\begin{eqnarray}
a_i^{\dagger}|\Omega\rangle \to |i\rangle:=(0,\cdots, \underbrace{1}_{i\text{-th position}},\cdots, 0);\quad i=\{1,2,\cdots,N\}
\end{eqnarray}
the Hamiltonian becomes an $N\times N$ matrix. However owing to the residual $\mathcal{S}_{N-1}$ symmetry, we further can reduce the dimension of the matrix. For example, we can work in the space spanned by 
\begin{eqnarray}
\bigg\{|1\rangle ,|2\rangle ,\frac{1}{\sqrt{N-2}}\left(|3\rangle + \cdots + |N\rangle \right)\bigg\}
\end{eqnarray}     
The resulting Hamiltonian is
\begin{eqnarray}
H=\begin{pmatrix}
0 & \beta & \beta \sqrt{N-2}\\
\beta & 0 & \sqrt{N-2}\\
\beta \sqrt{N-2} & \sqrt{N-2} & N-3\\
\end{pmatrix}
\end{eqnarray}
Using this the probability that after evolving $|1\rangle$ in time, we still find it at $|1\rangle$ comes to be
\begin{eqnarray}
|\langle 1|1(t)\rangle |^2=1-\frac{2(N-1)\beta^2\left(1-\cos{[t\sqrt{((N-2)^2+4(N-1)\beta^2)}]}\right)}{(N-2)^2+4(N-1)\beta^2}
\end{eqnarray}   
Similarly we can compute $|\langle 2|2(t)\rangle|^2$, and we show the plots for both of these in Figs. \ref{fig:p1}, \ref{fig:p2}. 

The essential difference between these two is that, we can localize $|1\rangle$ even for small $N$ by tuning $\beta$, which is not possible for the other states that continue to localise for large $N$ values as in the case of the full global $\mathcal{S}_N$ symmetry.

\paragraph{Models with reduced global symmetry - $\mathcal{S}_{N-k}$ :}
We expect similar statements to be true for a system with a global $\mathcal{S}_{N-k}\subset\mathcal{S}_N$ symmetry where $k$ points are marked. 
Consider the oscillators indexed by the conjugate variables $\alpha$,
\begin{equation}\label{eq:Atheta}
    A_\alpha = \frac{1}{\sqrt{N-k+\sum\limits_{j=1}^k|\theta_j|^2}}\left[\sum\limits_{j=1}^k~\omega^{\alpha j}\theta_j a_j + \sum\limits_{j=k+1}^N~\omega^{\alpha j} a_j \right],
\end{equation}
with $\theta_j$ being complex parameters. Using this, the Hamiltonian\footnote{Note that unlike the Hamiltonian for the $\mathcal{S}_{N-1}$ case, this Hamiltonian also includes the diagonal terms of the form, $a^\dag_ja_j$. Removing them amounts to including the term $\sum\limits_{\alpha=1}^N~A_\alpha^\dag A_\alpha$ to the Hamiltonian. However, unlike the undeformed case, this term does not commute with $A_N^\dag A_N$ when the oscillators are deformed as \eqref{eq:Atheta}. This only adds a layer of complication in the computation but we do not expect the features to alter much and so we do not include it here.}
\begin{equation}
    H = A_N^\dag A_N,
\end{equation}
is a simple example of a system with global $\mathcal{S}_{N-k}$ symmetry. In the original indices, this looks like
\begin{eqnarray}
\begin{split}
    H=\frac{1}{N-k+\sum\limits_{m=1}^k|\theta_m|^2}\Big[\sum\limits_{m,j=1}^k\theta_m^*\theta_ja^{\dagger}_m a_j+\sum\limits_{m,j=k+1}^N a^{\dagger}_m a_j+ \\ \sum\limits_{m=1}^k\sum\limits_{j=k+1}^N\left(\theta^*_ma^{\dagger}_m a_j+\theta_m a^{\dagger}_j a_m\right)\Big]
    \end{split}
\end{eqnarray}which explicitly reveals the $\mathcal{S}_{N-k}$ global symmetry among the indices $\{k+1,\cdots,N\}$. It is easily verified that oscillators \eqref{eq:Atheta} satisfy a deformed CAR algebra,
\begin{eqnarray}\label{eq:CARthetaSNk}
    \{A_\alpha, A_\beta\} & = & 0 = \{A_\alpha^\dag, A_\beta^\dag\}, \nonumber \\
    \{A_\alpha, A_\alpha^\dag\} & = & 1, \nonumber \\
    \{A_\alpha, A_\beta^\dag\} & = & \kappa_{\alpha\beta};~\alpha\neq\beta,
\end{eqnarray}
with $\kappa_{\alpha\beta} = \frac{\sum\limits_{j=1}^k~\omega^{(\alpha-\beta)j}|\theta_j|^2 + \sum\limits_{j=k+1}^N~\omega^{(\alpha-\beta)j}}{N-k+\sum\limits_{j=1}^k|\theta_j|^2}$. Note that when the deformation parameters $\theta_j\rightarrow 1$, this algebra reduces to the undeformed CAR algebra, \eqref{eq:CARalpha}. The spectrum of this system is similar to the fully symmetric case as each particle number sector has precisely two eigenvalues. The 1-fermion and 2-fermion spectrums are shown in Tables \ref{tab:1-fermionthetaNk}, \ref{tab:2-fermionthetaNk} to illustrate this.
\begin{table}[h!]
\centering
\begin{tabular}{ |c|c|c| } 
 \hline
 Eigenvalue & Eigenstate & Degeneracy \\
 \hline
 $1$ & $A_N^\dag\ket{\Omega}$ & 1 \\ 
 \hline
 0 & $\left(A_\alpha^\dag-\kappa_{N \alpha }A_N^\dag\right)\ket{\Omega};~\alpha=\{1,\cdots N-1\}$ & $N-1$ \\ 
 \hline
 \end{tabular}
 \caption{The 1-fermion spectrum of the system with reduced global symmetry, $\mathcal{S}_{N-k}$.}
 \label{tab:1-fermionthetaNk}
\end{table}

\begin{figure}[h!]
\centering
\begin{subfigure}{.5\textwidth}
  \centering
  \includegraphics[width=1\linewidth]{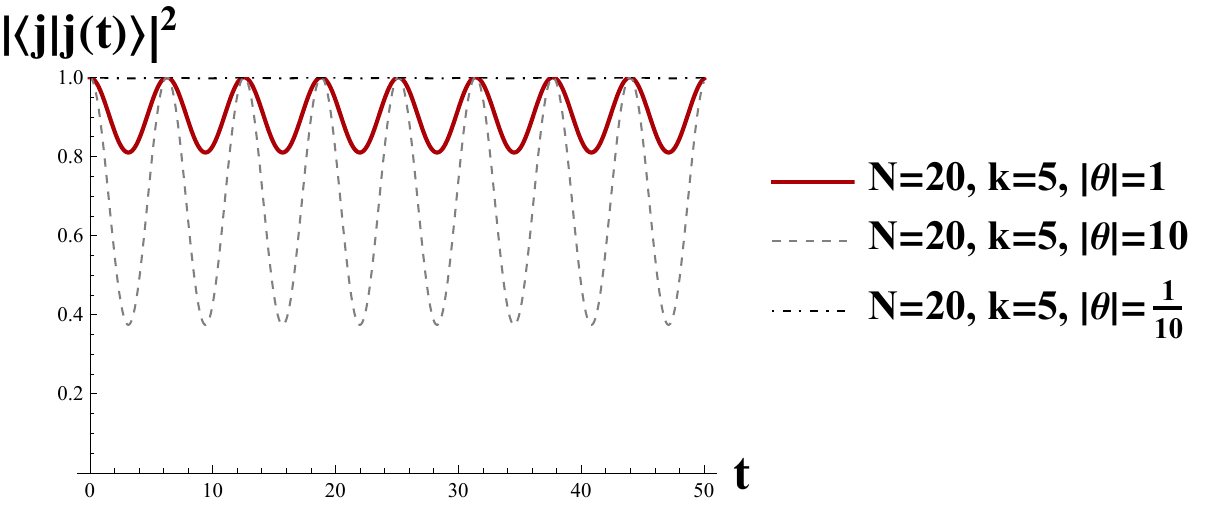}
  \caption{ }
  \label{fig:jjtSNktheta}
\end{subfigure}%
\begin{subfigure}{.5\textwidth}
  \centering
  \includegraphics[width=1\linewidth]{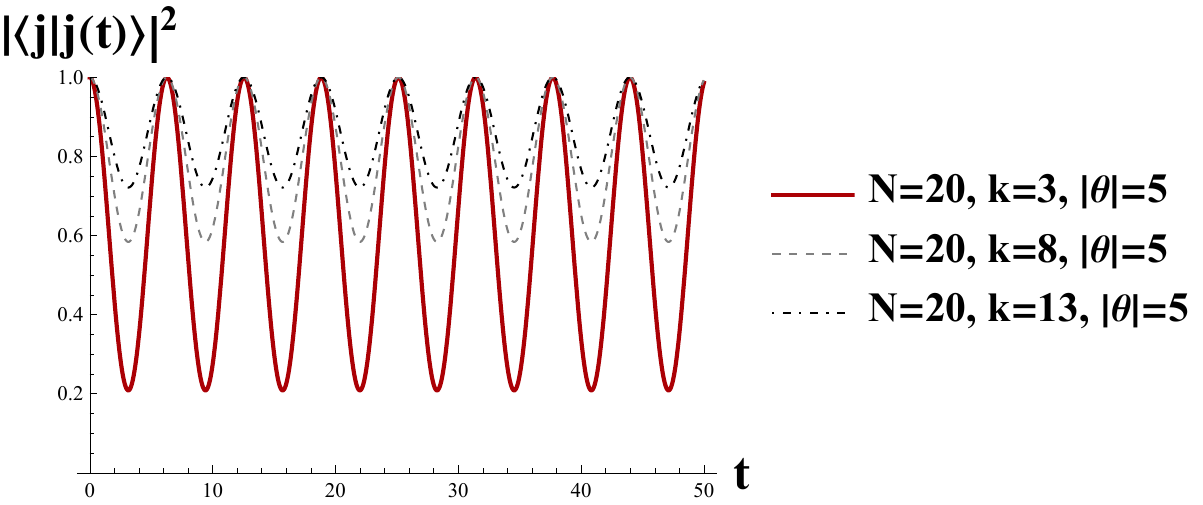}
  \caption{}
  \label{fig:jjtSNkk}
\end{subfigure}
\begin{subfigure}{.5\textwidth}
  \centering
  \includegraphics[width=1\linewidth]{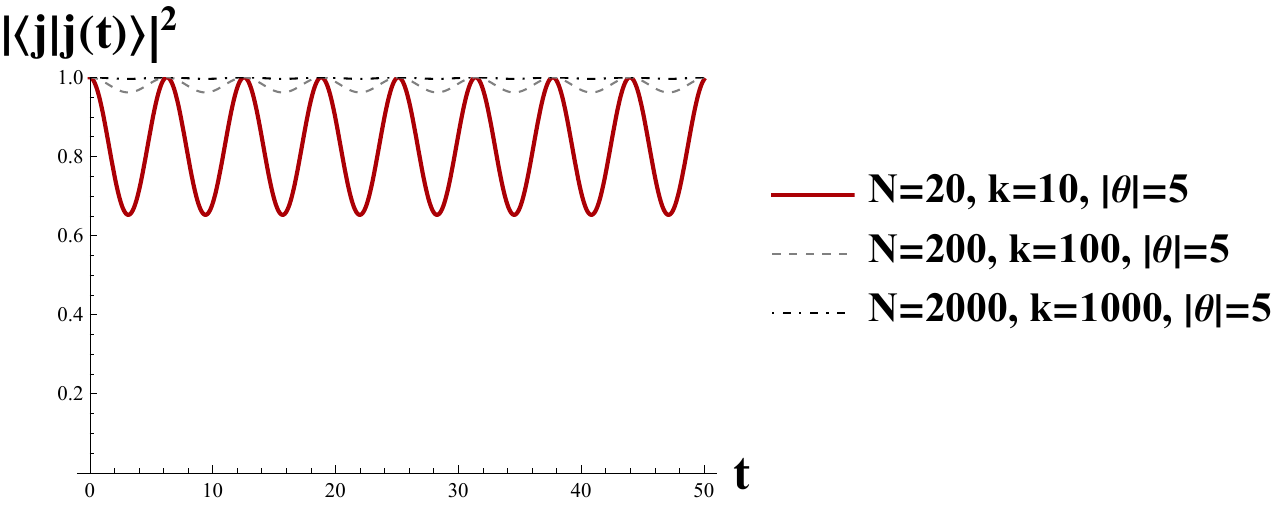}
  \caption{}
  \label{fig:jjtSNkkN}
\end{subfigure}
\caption{(Color Online) The probability distribution $|\langle j|j(t)\rangle|^2$, as given by (\ref{eq:jjtprobSNk}), is plotted against $t$ for different $\theta$ and $k$ values in $(a)$ and $(b)$ respectively. In $(c)$ we have plotted for different $k$ and $N$ values with fixed $|\theta|$ such that a particular $\frac{k}{N}$ ratio (here $0.5$) is maintained. One can arrive at both localization and delocalization by modifying $\theta$ and $k$ suitably just as in the $\mathcal{S}_{N-1}$ case studied earlier. Large values of $N$, with $\frac{k}{N}$ fixed and finite, favors localization as seen in $(c)$.}  
\end{figure}

\begin{table}[h!]
\centering
\begin{tabular}{ |c|c|c| } 
 \hline
 Eigenvalue & Eigenstate & Degeneracy \\
 \hline
 $1$ & $A_N^\dag A_\alpha^\dag\ket{\Omega};~\alpha\neq N$ & $N-1$ \\ 
 \hline
 $0$ & $\left(A_{\alpha_1}^\dag A_{\alpha_2}^\dag - \kappa_{N\alpha_1} A_N^\dag A_{\alpha_2}^\dag +  \kappa_{N\alpha_2} A_N^\dag A_{\alpha_1}^\dag\right)\ket{\Omega};~\alpha_1<\alpha_2\neq N$ & $\binom{N-1}{2}$ \\
 \hline
 \end{tabular}
 \caption{The 2-fermion spectrum of the system with reduced global symmetry, $\mathcal{S}_{N-k}$.}
 \label{tab:2-fermionthetaNk}
\end{table}
We observe that the value of $k$ and the deformation coefficients $\theta_j$ do not affect the nature of the spectrum, i.e. they continue to have precisely two eigenvalues for a Hamiltonian taking the form, $A_N^\dag A_N$. We expect the resulting probability distributions to depend on the values of $\theta_j$ just as the $k=1$ case studied earlier. For example when $j\in\{1,2,\cdots , k\}$ we have for the time evolved state,
\begin{eqnarray}\label{eq:jtSNk}
    \ket{j(t)} & = & \frac{1}{N}\left[N + \left(e^{-\mathrm{i}t} -1 \right)\sum\limits_{\alpha=1}^N~\omega^{j\alpha}\kappa_{N\alpha}\right]a_j^\dag\ket{\Omega} \nonumber \\
    & + & \frac{1}{N\theta_j^*}\mathop{\sum\limits_{l=1}^k}_{l\neq j}~\left(e^{-\mathrm{i}t} -1 \right)\sum\limits_{\alpha=1}^N~\omega^{j\alpha}\kappa_{N\alpha}\theta_l^* a_l^\dag\ket{\Omega} + \frac{1}{N\theta_j^*}\sum\limits_{l=k+1}^N~\left(e^{-\mathrm{i}t} -1 \right)\sum\limits_{\alpha=1}^N~\omega^{j\alpha}\kappa_{N\alpha} a_l^\dag\ket{\Omega} \nonumber \\
\end{eqnarray}
with normalisation $\mathcal{N}$ given by,
\begin{eqnarray}\label{eq:jtnormSNk}
    \mathcal{N}^2 & = & \frac{1}{N^2}\left[N + \left(e^{-\mathrm{i}t}-1 \right)\sum\limits_{\alpha_1=1}^N~\omega^{j\alpha_1}\kappa_{N\alpha_1} \right]\left[N + \left(e^{\mathrm{i}t}-1 \right)\sum\limits_{\alpha_2=1}^N~\omega^{-j\alpha_2}\kappa_{\alpha_2 N} \right] \nonumber \\
    & + & \frac{1}{N^2|\theta_j|^2}\left(e^{-\mathrm{i}t}-1 \right)\left(e^{\mathrm{i}t}-1 \right)\sum\limits_{\alpha_1=1}^N~\omega^{j\alpha_1}\kappa_{N\alpha_1}\sum\limits_{\alpha_2=1}^N~\omega^{-j\alpha_2}\kappa_{\alpha_2 N}\mathop{\sum\limits_{l=1}^k}_{l\neq j}~|\theta_l|^2 \nonumber \\
    & + &  \frac{N-k}{N^2|\theta_j|^2}\left(e^{-\mathrm{i}t}-1 \right)\left(e^{\mathrm{i}t}-1 \right)\sum\limits_{\alpha_1=1}^N~\omega^{j\alpha_1}\kappa_{N\alpha_1}\sum\limits_{\alpha_2=1}^N~\omega^{-j\alpha_2}\kappa_{\alpha_2 N}.
\end{eqnarray}
The apparent time-dependence in the above expression should vanish upon simplification as expected for a unitary system and we will see this explicitly below.
The system with arbitrary complex $\theta_j$'s is hard to analyse and so to better understand the behavior of the probability distributions arising from these expressions we make a simplifying assumption that $|\theta_j|^2=|\theta|^2>1$ for all $j\in\{1,2,\cdots, k \}$. This is still a fairly general consideration as the different $\theta_j$'s differ in the phases though they have the same magnitude. This simplification makes 
\begin{equation}
    \kappa_{N\alpha} = \frac{|\theta|^2-1}{N+k\left(|\theta|^2-1 \right)}\sum\limits_{j=1}^k~\omega^{-\alpha j} = \delta\sum\limits_{j=1}^k~\omega^{-\alpha j},
\end{equation} 
when $\alpha\neq N$. Using this we see that the frequently occurring sum simplifies as, 
\begin{equation}
    \sum\limits_{\alpha=1}^N~\omega^{\alpha j}\kappa_{N\alpha}=\begin{cases}
			1 + (N-k)\delta, & \text{if $j\in\{1,2,\cdots, k\}$ }\\
            1-k\delta & \text{if $j\in\{k+1,\cdots, N\}$}.
		 \end{cases}
\end{equation}
With this the normalisation in \eqref{eq:jtnormSNk} simplifies to one.
For $j\in\{1,2,\cdots, k\}$ the different probability distributions are found to be,
\begin{eqnarray}
    |\vev{j|j(t)}|^2 & = & \frac{\left[N + \left(e^{-\mathrm{i}t}-1 \right)\left(1+(N-k)\delta\right)  \right]\left[N + \left(e^{\mathrm{i}t}-1 \right)\left(1+(N-k)\delta\right)  \right]}{N^2}, \label{eq:jjtprobSNk}\\ 
     |\vev{l|j(t)}|^2 & = & \frac{\left(e^{-\mathrm{i}t}-1 \right)\left(e^{\mathrm{i}t}-1 \right)\left(1+(N-k)\delta\right)^2}{N^2};~~l\in\{1,2,\cdots, k \},~l\neq j, \label{eq:ljtprobSNk} \\
     |\vev{m|j(t)}|^2 & = & \frac{\left(e^{-\mathrm{i}t}-1 \right)\left(e^{\mathrm{i}t}-1 \right)\left(1+(N-k)\delta\right)^2}{N^2|\theta|^2};~~m\in\{k+1\cdots, N \}. \label{eq:mjtprobSNk}
\end{eqnarray}
It is easily seen that the sum of \eqref{eq:jjtprobSNk} and $k-1$ times \eqref{eq:ljtprobSNk} and $N-k$ times \eqref{eq:mjtprobSNk} gives one implying probability conservation and this serves as a consistency check on our expressions. From these expressions it is clear that for large enough $N$ when compared to $\theta$, only the expression in \eqref{eq:jjtprobSNk} goes to one while the expressions in \eqref{eq:ljtprobSNk} and \eqref{eq:mjtprobSNk} go to zero implying localisation. Furthermore, as in the $\mathcal{S}_{N-1}$ case, with the introduction of the deformation parameters $\theta$'s, we see that we can now achieve localisation by tuning their values for fixed $N$ and $k$. This is shown in Fig. \ref{fig:jjtSNktheta} where for $N=20$ we obtain almost perfect localisation when $\theta=0.1$ which should be contrasted with the $\theta=10$ case shown in the same figure.

Note that for $\theta=1$ the deformed CAR algebra in \eqref{eq:CARthetaSNk} reduces to the undeformed one in \eqref{eq:CARalpha}. However this is still not the fully symmetric case due to the continued presence of the phases in front of the $k$ oscillators. Nevertheless the probability distributions are blind to these relative phases and hence the answers we obtain in this case are precisely the same as the undeformed or the fully symmetric case. This suggests that the fully deformed model, devoid of any global symmetry and built using the oscillators 
\begin{equation}
    A_\alpha = \frac{1}{|\theta|\sqrt{N}}\sum\limits_{j=1}^N~\omega^{\alpha j}\theta_ja_j,
\end{equation}
should mimic the undeformed and fully symmetric case. It is easily seen that the deformed CAR algebra, \eqref{eq:CARthetaSNk} reduces to the undeformed one, \eqref{eq:CARalpha} for these oscillators. In fact the probability distributions in this case,
\begin{eqnarray}
    |\vev{j|j(t)}|^2 & = & \frac{1}{N^2}\left[N + e^{\mathrm{i}t} -1 \right]\left[N + e^{-\mathrm{i}t} -1 \right], \\
     |\vev{l|j(t)}|^2 & = & \frac{1}{N^2}\left[e^{\mathrm{i}t} -1 \right]\left[e^{-\mathrm{i}t} -1 \right];~~l\neq j,
\end{eqnarray}
coincide precisely with the undeformed case. These expressions can also be obtained by setting $|\theta|=1$ or $\delta=0$ and $k=N$ in \eqref{eq:jjtprobSNk}, \eqref{eq:ljtprobSNk}. This is another example where a model without global symmetries localises in exactly the same way as a model with one, as long as all the fermions interact with each other modulo relative phase coefficients. 

The role of $k$ in $\mathcal{S}_{N-k}$ is studied in the plot \ref{fig:jjtSNkk} where we see that for fixed $N$ and $|\theta|$ the system localises as $k\rightarrow N$. Thus the tuning of $k$ can also be used to obtain localisation for small $N$. Moreover as we increase $k$ and $N$ such that $\frac{k}{N}$ is fixed we see that localisation occurs only for large $N$ when $|\theta|$ is fixed (See Fig. \ref{fig:jjtSNkkN}). In this case localisation for smaller values of $N$ can be obtained by modulating $|\theta|$. Also when $\frac{k}{N}\rightarrow 1$ we will always see localisation. 

As a further extension we remark that Hamiltonians of the form $A_\alpha^\dag A_\alpha$, diagonal in the conjugate space, will have similar spectral features as the fully symmetric case, for both deformed and non-deformed oscillators. These systems do not have any obvious global symmetries in the spaces indices $j$ but feature interactions where all the fermions interact with one another, with some coefficients. We expect all of them to exhibit similar disorder-free localisation features.

\paragraph{Spin chain with global $\mathcal{S}_N$ symmetry :} To examine the role of the superselected symmetry, consider a spin chain system with no superselected symmetry and just a global $\mathcal{S}_N$ symmetry. 

The permutation operator on $\mathbb{C}^2\otimes\mathbb{C}^2$ can be written using Pauli matrices as
\begin{equation}
    \frac{1+X\otimes X + Y\otimes Y + Z\otimes Z}{2}=\begin{pmatrix}
    1 & 0 & 0 & 0 \\
    0 & 0 & 1 & 0 \\
    0 & 1 & 0 & 0 \\
    0 & 0 & 0 & 1 \\
\end{pmatrix}.
\end{equation} 
Thus we can generate the full permutation group $\mathcal{S}_N$ using these transpositions as generators. We have
\begin{equation}\label{eq:SNspinrep}
    s_i\equiv s_{i, i+1}=\frac{1+X_iX_{i+1} + Y_iY_{i+1} + Z_iZ_{i+1}}{2}
\end{equation}
as the generators of $\mathcal{S}_N$ and they satisfy \eqref{eq:SNrelations}. Well-known spin chains such as the Heisenberg $XXX$ spin chain, $H=\sum\limits_{j=1}^{N-1}~\left[X_jX_{j+1} + Y_jY_{j+1} + Z_jZ_{j+1}\right]$ is a sum of such operators up to a constant term. For our purposes we use these generators to construct permutation invariant spin chains using the techniques of Sec.\ref{sec:construction}. By noting that the generators are just transpositions or 2-cycles, we can express an arbitrary 2-cycle, $(ij)$ as a product of the $s_i$'s in \eqref{eq:SNspinrep}. We have,  
\begin{equation}
    (ij) = \frac{1+X_iX_j + Y_iY_j + Z_iZ_j}{2},
\end{equation}
where $X_i$, $Y_i$ and $Z_i$ are the Pauli matrices on site $i$. This operator exchanges the states on the sites $i$ and $j$. In this case the 2-cycle Hamiltonian generalizes the Heisenberg $XXX$ spin chain\footnote{The Heisenberg $XXX$ spin chain, $H=\sum\limits_{j=1}^{N-1}~\left[X_jX_{j+1} + Y_jY_{j+1} + Z_jZ_{j+1}\right]$ is a local Hamiltonian with nearest-neighbor interactions. This can be thought of as the analogue of the tight-binding Hamiltonian. The analogue of the $\mathcal{S}_N$-symmetric Hamiltonian in \eqref{eq:2-cycleH-alltoall} is the model given by \eqref{eq:symmetricXXX}.},
\begin{equation}\label{eq:symmetricXXX}
   H = \frac{1}{2}\mathop{\sum\limits_{j,k=1}^N}_{j<k}~ \left[X_jX_k + Y_jY_k + Z_jZ_k\right] + \frac{N(N-1)}{2}.
\end{equation}
The 2-cycle Hamiltonian is just the sum of the $\frac{N(N-1)}{2}$ transpositions acting on $\otimes_{j=1}^N\mathbb{C}^2_j$. We will call this the {\it symmetric Heisenberg $XXX$ spin chain.} 

The Hilbert space of this system splits into sectors labelled by the number of $\ket{\downarrow}$'s as the Hamiltonian, being the sum of permutation operators, cannot mix states containing different number of $\ket{\downarrow}$'s. Thus the problem we have now is similar to the one encountered in the fermionic realisation. For example consider the sector where there is a single $\ket{\downarrow}$ with the remaining sites filled by a $\ket{\uparrow}$. There are $N$ such states and the Hamiltonian in this sector reduces to the form,
\begin{equation}
    (H_1)_{ij} = \left(\frac{(N-1)(N-2)}{2} -1\right)\delta_{ij} + 1 
\end{equation}
which is similar to the one obtained in the fermion case, (Appendix \ref{eq:Halltoall-1fermion}) modulo the function of $N$ appearing with the $\delta_{ij}$. We expect to see localisation here as well suggesting that this property is true regardless of the realisation chosen for the global permutation symmetry. It is not hard to see that this pattern continues for the other sectors of this Hilbert space and so we conclude that the symmetric $XXX$ chain of \eqref{eq:symmetricXXX} will contain localised states for large $N$.

\paragraph{Bosonic systems with global $\mathcal{S}_N$ symmetry :} The symmetric Hamiltonian in \eqref{eq:2-cycleH-alltoall} can also be considered for the bosonic case by merely enforcing the canonical commutation relations (CCR) for the operators appearing in the Hamiltonian. While the 1-particle spectrum of this model is similar to the fermionic case, the spectrum of other sectors are different. For the higher particle sectors the number of energy eigenvalues depends on the particle number of the sector. There are $k+1$ energy eigenvalues for the $k$-particle sector. The spectrum of the 2-particle sector is shown in Table \ref{tab:2-particleBoson}. 
\begin{table}[h!]
\centering
\begin{tabular}{ |c|c|c| } 
 \hline
 Eigenvalue & Eigenstate & Degeneracy \\
 \hline
 $-2$ & $A_\alpha^\dag A_\beta^\dag\ket{\Omega};~\alpha,\beta\neq N$ & $\binom{N-1}{2}+N-1$ \\ 
 \hline
 $2N-2$ & $(A_N^\dag)^2\ket{\Omega}$ & $1$ \\ 
 \hline
 $N-2$ & $A_\alpha^\dag A_N^\dag\ket{\Omega}$ & $N-1$ \\
 \hline
 \end{tabular}
 \caption{The 2-particle spectrum of \eqref{eq:2-cycleH-alltoall} in the bosonic case.}
 \label{tab:2-particleBoson}
\end{table}
The probability for the 2-particle basis states is found to be,
\begin{eqnarray}
    |\vev{1,2|1,2(t)}|^2 & = & \frac{2}{N^4}\left[\frac{1+4(N-1)^2+(N-1)^4}{2} + 2(N-1)\cos{(E_1-E_2)t} \right. \nonumber \\
    & + & \left. 2(N-1)^3\cos{(E_2-E_3)t} + (N-1)^2\cos{(E_3-E_1)t}\right].
\end{eqnarray}
For large $N$ this expression clearly goes to unity indicating the same type of disorder-free localisation as seen in the fermionic system. The probabilities go as $\frac{1}{N^4}$ and $\frac{1+(N-2)^2+(N-1)^2}{N^4}$ when $\ket{1,2(t)}$ overlaps with $\ket{j,k}$ and $\ket{1,k}, \ket{2,k}$ respectively. Similar localisation is seen in the higher boson sectors as well and thus we conclude that this type of disorder-free localisation is also true for bosonic systems with global $\mathcal{S}_N$ symmetry.  

\subsection{A prescription for disorder-free localisation using graph theory}
\label{subsec:DFLgraphs}
We begin with a few definitions from graph theory to keep this part self-contained. A graph $G=(V,E)$ consists of a set of vertices $(V)$ and a set of edges $(E)$. We consider simple and regular graphs which are those where more than one edge is not allowed between two vertices and where all the vertices have the same valency. The adjacency matrix $A$ of a simple and regular graph is a $|V|\times |V|$ matrix with matrix elements $A_{ij} = 1$ when vertices $i$ and $j$ are connected by an edge and 0 when they are not. 

From the graph theory perspective the Hamiltonian of the 1-fermion sector (\eqref{eq:Halltoall-1fermion}) is precisely the {\it adjacency matrix} of a complete graph\footnote{A complete graph is a simple graph where all vertices are connected with each other.}. The automorphism group of such a graph is precisely $\mathcal{S}_N$, where $N$ is identified with the number of vertices of this graph.

Now we establish the localisation features seen in the 1-fermion Hamiltonian (\eqref{eq:Halltoall-1fermion}) for the complete graph. To do this we begin by making the set of vertices into a Hilbert space, $\mathbb{C}^N$ with the $N$ basis vectors given by $\ket{j} = \left(
    0  \cdots  1  \cdots  0\right)^T$, with 1 at the $j$th position, and the inner product taken as the canonical one. Then the vectors $\ket{j}$ are evolved according to 
\begin{equation}
    \ket{j(t)} = e^{\mathrm{i}At}\ket{j}.
\end{equation}
Computing the probability corresponding to the overlap $\vev{1|1(t)}$,
\begin{equation}
    |\vev{1|1(t)}|^2 = \frac{1}{N^2}\left[1+(N-1)^2 + 2(N-1)\cos{Nt} \right],
\end{equation}
we see that it coincides with the 1-fermion quantum walk expression, \eqref{eq:1-fermionprobdistr1}. The probability distributions corresponding to states other than $\ket{1}$ precisely coincide with \eqref{eq:1-fermionprobdistrj}. These results have appeared in the quantum walk literature from the perspective of algebraic graph theory (See Sec. 1 of \cite{Godsil2017SedentaryQW}, Eq. 15 of \cite{Xu2009ExactAR}, Eq. 4 of \cite{Frigerio2022SwiftCQ}, and \cite{Godsil2010WhenCP}, \cite{Ide2014LocalSS}). Such localisation properties appear under the guise of `sedentariness in continuous time quantum walks' and the `stay-at-home' property in the graph theory jargon. More generally graph theorists study {\it strongly regular graphs} (SRG) \cite{Godsil2001AlgebraicGT, Brouwer2011SpectraOG} whose adjacency matrices have spectra with the right properties to obtain the stay-at-home property \cite{Godsil2017SedentaryQW}. 
SRG's are characterized by a 4-tuple, $\left(n,k;a,c\right)$ with $n$ being the number of vertices, $k$ the valency, $a$ the number of common neighbors for adjacent vertices and $c$ the number of common neighbors for non-adjacent vertices. Their adjacency matrices satisfy the characteristic equation \cite{Godsil2001AlgebraicGT, Godsil2017SedentaryQW},
\begin{equation}
    A^2 - (a-c)A -(k-c) I = cJ,
\end{equation}
where I is the identity matrix of size $n$ and $J$ is the square matrix with all entries unity. This matrix has just three eigenvalues, one of which is non-degenerate and the other two scale with $n$. 

Not all strongly regular graphs exhibit localisation but a particular family characterised by the tuple $\left(n^2, k(n-1); n-2 + (k-1)(k-2), k(k-1)\right)$ does show localisation \cite{Godsil2017SedentaryQW}. Most of the strongly regular graphs have trivial automorphism groups \cite{Godsil2001AlgebraicGT} which shows that the global symmetry is really not necessary to see this kind of disorder-free localisation.  

Interpreting the adjacency matrices as many-body Hamiltonians we see that they only capture the 1-particle sectors of the corresponding physical systems. In this paper we have shown how to construct the physical Hamiltonians\footnote{Note that in graph theory the localisation results are derived purely by studying the characteristic equation which translates to the spectra of the underlying graphs. They do not require the exact form of the adjacency matrix to arrive at their results. In fact it is a hard problem to construct the adjacency matrices of strongly regular graphs \cite{Godsil2001AlgebraicGT}.} corresponding to such graphs which facilitate the study of the localisation properties of multi-particle sectors as well. For example the fermionic Hamiltonian corresponding to an adjacency matrix $A$ can be constructed by replacing the matrix elements of $A$ by $a^\dag_ia_j + a^\dag_ja_i$ iff $A_{ij}=1$. With this rule the fermionic Hamiltonian corresponding to the complete graph is precisely the $\mathcal{S}_N$-symmetric model in \eqref{eq:2-cycleH-alltoall}. This can then be used to study the higher fermion sectors. With this algorithm it is only natural to conjecture that the Hamiltonians corresponding to SRG's can be prescribed as a way to obtain the disorder-free localisation studied in this paper. We reserve the investigation of this interesting connection for a future work.

\subsection{Final remarks}
\label{subsec:finalremarks}

We end with a few remarks and possible future works.
\begin{enumerate}
   \item The models described in this paper feature all-to-all connectivity of multiple fermions which are essentially two-level systems implying that they can be realised with qubit systems. The many-fermion Hamiltonians in \eqref{eq:2-cycleH-alltoall}, \eqref{eq:Htheta} and \eqref{eq:quartic H} can be modelled on both planar and non-planar architectures depending on the potentials used to realise the inter-qubit couplings. However it is in general impractical and hard to engineer such long-range couplings between qubits in experiments. Nevertheless such systems are important in quantum computing architectures as they can implement different quantum algorithms that assume quantum gates operate on an arbitrary pair of qubits. An implementation of the homogeneous all-to-all connectivity among qubits is proposed in \cite{tsomokos2008fully} using superconducting qubits in circuit QED. Also there are now experiments in superconducting circuits using bus resonators that can achieve tunable all-to-all couplings \cite{Xu2019ProbingTD, Song2019GenerationOM} and more recently with ring resonators \cite{Hazra2020LongrangeCI}. The architecture proposed in the latter is also scalable and prevents cross-talk between qubits. 
   
  All-to-all connectivity can also be seen in digital quantum simulators using trapped ions \cite{Lanyon2011UniversalDQ, Wright2019BenchmarkingA1}. In these setups the qubits are encoded in the Zeeman states of electrically trapped and laser-cooled calcium ions. A universal set of gates are implemented on this system and they can simulate several types of interactions among the qubits including non-local ones apart from the local terms. These experiments also support both inhomogeneous and homogeneous couplings. Similar effects are also seen in ultracold atom setups \cite{Cazalilla2014UltracoldFG}.
    
  \item The {\it out-of-time-order correlators} (OTOC's) indicate chaotic behavior in thermal systems \cite{Hashimoto2017OutoftimeorderCI}. The Lyapunov exponent can be read off from such expressions \cite{Maldacena2015ABO}. The saturation of this quantity implies behavior similar to that predicted by the AdS/CFT correspondence and is like an SYK model \cite{KitaevSeminar,Sachdev1992GaplessSG,Maldacena2015ABO,Rosenhaus2018AnIT}. The systems discussed in this work describe the opposite behavior and thus the results on the OTOC's do not hold for them.
    \item An extension worthy of mention is adding an internal symmetry index $\alpha=1,2$ transforming by the spin $\frac{1}{2}$ representation of $SU(2)$ to the operators $a_i$. Then since this representation is pseudoreal, $a_i^\alpha\left(\sigma_2\right)_{\alpha, \alpha'}a_j^{\alpha'}$ is $SU(2)$ invariant and so is its adjoint (Here $\sigma_2$ is the second Pauli matrix.). So we can add such $SU(2)$ or colour singlet Majorana terms which are also permutation invariant and incorporate features of the SYK model.
    \item Furthermore it would be interesting to consider the algebra of observables that are permutation invariant and study the corresponding Hilbert spaces built using the GNS construction \cite{Balachandran2013EntanglementAP, Balachandran2013AlgebraicAT,  Reyes-Lega:2022sbq, Balachandran:2012pa, Balachandran:2013kia, Benatti2020EntanglementII}. These can then be used to explore the entanglement entropy and thermalisation properties of these systems and those derived from them.
\end{enumerate}
 
\section*{Acknowledgments}
Certain ideas in this paper were initiated in discussions of A.P.B and Fabio Di Cosmo which we gratefully acknowledge. A.P.B enjoyed the hospitality of The Institute of Mathematical Sciences, Chennai while this work was in progress. For that, he is grateful to the Director, Ravindran, and his colleague and friend Sanatan Digal. AK acknowledges support of the Department of Atomic Energy, Government of India, under project no. RTI4001. A.S and P.P thank Tapan Mishra and Abhishek Chowdhury for useful discussions. A.K and P.P acknowledge useful comments and suggestions from Sibasish Ghosh. P.P thanks G.Baskaran and Ayan Mukhopadhyay for useful discussions and the latter also for the hospitality at IIT Madras. P.P also acknowledges discussions with Kristian Hauser Villegas and Kim Kun Woo. We would also like to thank the anonymous referees of Physical Review A for several critical comments that helped improve the paper. 

\appendix
\section{$H^{(2)}_1$ in different particle sectors}
\label{app:H1 every sector}
Let us consider the operator
\begin{eqnarray}
H^{(2)}_1=\sum_{\substack{\rho\in 2\text{-cycle}\\ \text{conjugacy class}}}\sum_{i=1}^N a^{\dagger}_{\rho(i)}a_i
\end{eqnarray}
This is invariant in every particle sector under global permutations among the site indices. In one-particle sector this can be seen very easily :
\begin{eqnarray}
\left(\sigma_1 H^{(2)}_1 \sigma^{-1}_1\right)a_m^{\dagger}|\Omega\rangle=H^{(2)}_1a_m^{\dagger}|\Omega\rangle
\end{eqnarray}
Here $\sigma_1$ is the realization of  $\sigma\in\mathcal{S}_N$ in the one-particle sector.

However before proceeding further let us consider the action of $H^{(2)}_1$ on an arbitrary $k$-particle state, $a^{\dagger}_{i_1}\cdots a^{\dagger}_{i_k}|\Omega\rangle$. We can write
\begin{eqnarray}
H^{(2)}_1a^{\dagger}_{i_1}\cdots a^{\dagger}_{i_k}|\Omega\rangle &=& \sum_{\substack{\rho\in 2\text{-cycle}\\ \text{conjugacy class}}}\left(a^{\dagger}_{\rho(i_1)}a_{i_1}+\cdots a^{\dagger}_{\rho(m)}a_m+\cdots +a^{\dagger}_{\rho(i_k)}a_{i_k}\right)a^{\dagger}_{i_1}\cdots a^{\dagger}_{i_k}|\Omega\rangle\quad\quad\text{(rests give $0$)}\nonumber\\
&=&\sum_{\substack{\rho\in 2\text{-cycle}\\ \text{conjugacy class}}}\left(a^{\dagger}_{\rho(i_1)}\cdots a^{\dagger}_{i_k}+a^{\dagger}_{i_1}\cdots a^{\dagger}_{\rho(m)}\cdots a^{\dagger}_{i_k}+a^{\dagger}_{i_1}\cdots a^{\dagger}_{\rho(i_k)}\right)|\Omega\rangle 
\end{eqnarray}
We will now demonstrate the invariance of $H^{(2)}_1$ in the $k$-particle sector.
\begin{eqnarray}
\sigma_kH^{(2)}_1 \sigma_k^{-1}|i_1,\cdots, i_k\rangle &=& \sigma_k H^{(2)}_1|\sigma^{-1}(i_1),\cdots,\sigma^{-1}(i_k)\rangle \nonumber \\
&=& \sigma_k\sum_{\substack{\rho\in 2\text{-cycle}\\ \text{conjugacy class}}}\left(|\rho\sigma ^{-1}(i_1),\cdots,\sigma^{-1}(i_k)\rangle +\cdots+|\sigma^{-1}(i_1),\cdots,\rho\sigma^{-1}(i_k)\rangle\right)\nonumber\\
&=&\sum_{\substack{\rho\in 2\text{-cycle}\\ \text{conjugacy class}}}\left(|\sigma\rho\sigma ^{-1}(i_1),\cdots,i_k\rangle +\cdots+|i_1,\cdots,\sigma\rho\sigma^{-1}(i_k)\rangle\right)\nonumber\\
&=&\sum_{\substack{\rho'\in 2\text{-cycle}\\ \text{conjugacy class}}}\left(|\rho '(i_1),\cdots,i_k\rangle +\cdots+|i_1,\cdots,\rho '(i_k)\rangle\right)\nonumber\\
&=& H^{(2)}_1|i_1,\cdots, i_k\rangle
\end{eqnarray}
where $\sigma \rho \sigma^{-1}=\rho'\in 2\text{-cycle conjugacy class}$ also and $\sigma_k$ is the realization of $\sigma\in\mathcal{S}_N$ in the $k$-particle sector.

\section{Proof of \eqref{eq:2-cycleH-2fermion}}
\label{app:2-cycleH-2fermion}
The quartic Hamiltonian originating from the two-cycle conjugacy class is
\begin{eqnarray}\label{quartic}
H_2^{(2)}=\sum_{\sigma}\sum_{i<j} a^{\dagger}_{\sigma(i)}a^{\dagger}_{\sigma(j)}a_ja_i=\frac{1}{2}\sum_{i,j}\sum_{\sigma} a^{\dagger}_{\sigma(i)}a^{\dagger}_{\sigma(j)}a_ja_i
\end{eqnarray} 
 We notice that all $\sigma $'s belonging to the two-cycle conjugacy class can be grouped as
 
$\bullet$ $i\to i\quad j\to j$\quad\quad there are $\;^{N-2}C_2$ such elements. 

$\bullet$ $i\to j\quad j\to i$\quad\quad there is one such element.

$\bullet$ $i\to i\quad j\to m(\neq i,j)$\quad\quad there are $N-2$ such elements.

$\bullet$ $i\to m(\neq i,j)\quad j\to j $\quad\quad there are $N-2$ such elements.

Considering this, we can simplify (\ref{quartic}) as
\begin{eqnarray}
H_2^{(2)}& = &\frac{1}{2}\sum_{i,j}\left[\;^{N-2}C_2 a^{\dagger}_ia^{\dagger}_ja_j a_i+a^{\dagger}_ja^{\dagger}_ia_ja_i+\sum_{m\neq i,j}\left(a^{\dagger}_ia^{\dagger}_m+a^{\dagger}_m a^{\dagger}_j\right)a_j a_i\right]
\end{eqnarray}
With $A=\sum_{i}a_i$, $\hat{N}=\sum_i a_i^{\dagger}a_i$ being the number operator and using the relations $[\hat{N},A/a]=-A/a$, we obtain
\begin{eqnarray}
H_2^{(2)}&=&\frac{1}{2} \left( ^{N-2}C_2-3\right)\left(\hat{N}^2-\hat{N}\right)+\left(A^{\dagger}A\hat{N}-A^{\dagger}A\right)\nonumber \\
&=&\frac{1}{2}\left(\frac{(N-2)(N-3)}{2}-1 \right)\left(\hat{N}^2-\hat{N} \right) + 2\sum\limits_{i<j}~ \left[a_i^\dag a_j + a_j^\dag a_i\right]\left(\hat{N}-1 \right)
\end{eqnarray}
So, we can write $H^{(2)}_2$ in terms of $H^{(2)}_1$ and $\hat{N}$ and we expect this trend to continue for higher order Hamiltonians like hextic Hamiltonian and so on.

\section{Orthogonality and Completeness of \eqref{eq:2-fermionji}, \eqref{eq:2-fermionjkN}}
\label{app:Completeness}
The inner product between 2-fermion eigenstates $\vev{u|v}$ is zero when $\ket{u}\in$ \eqref{eq:2-fermionji} and $\ket{v}\in$ \eqref{eq:2-fermionjkN}. However when both $\ket{u}$ and $\ket{v}$ belong to any particular eigenvalue then they may not be orthogonal to each other. Nevertheless we can show that these states are complete using the action of the permutation operators from $\mathcal{S}_N$. To do this we first note that the states in \eqref{eq:2-fermionji} are mapped into each other,
\begin{equation}
    s_{jk}\ket{[j]}_{2} = \ket{[k]}_{2}~;~j,k\in\{1,2,\cdots, N-1\},
\end{equation}
using the transpositions $s_{jk}$ and the states in \eqref{eq:2-fermionjkN} are mapped into each other,
\begin{equation}
    s_{j_1j_2}s_{k_1k_2}\ket{[j_1k_1N]}_{2} = \ket{[j_2k_2N]}_{2}
\end{equation}
using the permutations $s_{j_1j_2}s_{k_1k_2}$. Combining these identities with the expression of the 2-fermion state $\ket{1,2}$ in \eqref{eq:2-fermion12} we see that any other 2-fermion state $\ket{j,k}$, with $j,k\neq\{1,2\}$, can be obtained as 
\begin{eqnarray}
    \ket{j,k} & = &  s_{1j}s_{2k}\ket{1,2} \nonumber \\
              & = & \frac{\sqrt{N-1}}{N}\Big(\ket{[j]}_{2}-\ket{[k]}_{2}\Big) + \frac{\sqrt{3}(N-2)}{N}\ket{[jkN]}_{2} - \frac{\sqrt{3}}{N}\sum\limits_{m=3}^{N-1}\Big(\ket{[jmN]}_{2} - \ket{[kmN]}_{2}\Big). \nonumber \\
\end{eqnarray}
Notice that we have used $s_{1j}\ket{[2]}_{2}=\ket{[2]}_{2}$ and $s_{2k}\ket{[1]}_{2} = \ket{[1]}_{2}$. On the other hand the states of the form $\ket{1,j}$ ($\ket{2,j}$), for $j\in\{3,\cdots N\}$,  are obtained by applying $s_{2j}$ ($s_{1j}$) on $\pm\ket{1,2}$ respectively. Thus any 2-fermion state can be written as a linear combination of the 2-fermion eigenstates in \eqref{eq:2-fermionji} and \eqref{eq:2-fermionjkN} showing their completeness. These arguments can be extended to a general $k$-fermion sector as well.

\section{Proof of \eqref{eq:k-fermionstates2set}}
\label{app:k-fermionstates2set}
We have the action of quadratic all-to-all on a general $k$-particle state as
\begin{eqnarray}
H^{(2)}_1a^{\dagger}_{i_1}\cdots a^{\dagger}_{i_k}|\Omega\rangle &=&\sum_{\sigma}\left(a^{\dagger}_{\sigma(i_1)}\cdots a^{\dagger}_{i_k}+a^{\dagger}_{i_1}\cdots a^{\dagger}_{\sigma(i_j)}\cdots a^{\dagger}_{i_k}+\cdots +a^{\dagger}_{i_1}\cdots a^{\dagger}_{\sigma(i_k)}\right)|\Omega\rangle 
\end{eqnarray}
We concentrate on the single term
\begin{eqnarray}
\sum_{\sigma}a^{\dagger}_{i_1}\cdots a^{\dagger}_{\sigma(i_j)}\cdots a^{\dagger}_{i_k}
\end{eqnarray}
As done previously, we can group the $\sigma$'s as

$\bullet$ $i_j \to i_j$\quad\quad\text{there are $^{N-1}C_2$ such elements.}

$\bullet$ $i_j \to m\neq i_j$ \quad\quad\text{there are $N-1$ such elements.}

Then we finally have
\begin{eqnarray}
\sum_{\sigma}a^{\dagger}_{i_1}\cdots a^{\dagger}_{\sigma(i_j)}\cdots a^{\dagger}_{i_k}&=&^{N-1}C_2\;a^{\dagger}_{i_1}\cdots a^{\dagger}_{i_j}\cdots a^{\dagger}_{i_k}+a^{\dagger}_{i_1}\cdots \left(\sum_{m\neq i_j}a^{\dagger}_{m}\right)\cdots a^{\dagger}_{i_k}\nonumber \\
&=&\left(^{N-1}C_2-1\right)\;a^{\dagger}_{i_1}\cdots a^{\dagger}_{i_j}\cdots a^{\dagger}_{i_k}+a^{\dagger}_{i_1}\cdots \left(\sum_{m}a^{\dagger}_{m}\right)\cdots a^{\dagger}_{i_k}\nonumber \\
&=&\left(^NC_2-N\right)a^{\dagger}_{i_1}\cdots a^{\dagger}_{i_k}+a^{\dagger}_{i_1}\cdots \underbrace{A^{\dagger}}_{j\text{-th position}}\cdots a^{\dagger}_{i_k}
\end{eqnarray}
Therefore we have 
\begin{eqnarray}
H^{(2)}_1a^{\dagger}_{i_1}\cdots a^{\dagger}_{i_k}|\Omega\rangle &=& k\left(^NC_2-N\right)a^{\dagger}_{i_1}\cdots a^{\dagger}_{i_k}|\Omega\rangle + A^{\dagger}a^{\dagger}_{i_2}\cdots a^{\dagger}_{i_k}|\Omega\rangle + \cdots  +  a^{\dagger}_{i_1}\cdots \underbrace{A^{\dagger}}_{j\text{-th position}}\cdots a^{\dagger}_{i_k}|\Omega\rangle\nonumber \\ &&+ \cdots +  a^{\dagger}_{i_1}\cdots a^{\dagger}_{i_{k-1}}A^{\dagger}|\Omega\rangle
\end{eqnarray}
Now let us consider the following situation :
\begin{eqnarray}
H^{(2)}_1a^{\dagger}_{i_1}\cdots \sum_{i_k}a^{\dagger}_{i_k}|\Omega\rangle &=& k\left(^NC_2-N\right)a^{\dagger}_{i_1}\cdots \sum_{i_k}a^{\dagger}_{i_k}|\Omega\rangle + A^{\dagger}a^{\dagger}_{i_2}\cdots \sum_{i_k}a^{\dagger}_{i_k}|\Omega\rangle +
\nonumber \\ 
&& \cdots  +  a^{\dagger}_{i_1}\cdots \underbrace{A^{\dagger}}_{j\text{-th position}}\cdots \sum_{i_k}a^{\dagger}_{i_k}|\Omega\rangle + \cdots +  a^{\dagger}_{i_1}\cdots a^{\dagger}_{i_{k-1}}\sum_{i_k}A^{\dagger}|\Omega\rangle\nonumber\\
&=& k\left(^NC_2-N\right)a^{\dagger}_{i_1}\cdots A^{\dagger}|\Omega\rangle + N a^{\dagger}_{i_1}\cdots A^{\dagger}|\Omega\rangle 
\end{eqnarray}
Thus we have the following eigenstates of $H^{(2)}_1$:
\begin{eqnarray}
H^{(2)}_1a^{\dagger}_{i_1}\cdots A^{\dagger}|\Omega\rangle = \left(k ^N C_2-N(k-1)\right)a^{\dagger}_{i_1}\cdots A^{\dagger}|\Omega\rangle
\end{eqnarray}
Now let us consider another kind of states 
\begin{eqnarray}
|i_1,\cdots, i_k,N;k\rangle &=& a^{\dagger}_{i_1}\cdots a^{\dagger}_{i_k}|\Omega\rangle + \cdots +(-1)^{j k} a^{\dagger}_{j+1}\cdots a^{\dagger}_{i_k} \underbrace{a^{\dagger}_{N}}_{k-j+1\text{-th position}}a^{\dagger}_{i_1}\cdots a^{\dagger}_{j-1}|\Omega\rangle + \nonumber \\ 
&&\cdots + (-1)^{k^2}a^{\dagger}_N a^{\dagger}_{i_1}\cdots a^{\dagger}_{i_{k-1}}|\Omega\rangle 
\end{eqnarray}
Let us focus on two particular expressions
\begin{eqnarray}
H^{(2)}_1a^{\dagger}_{i_1}\cdots a^{\dagger}_{i_k}|\Omega\rangle\quad\text{and}\quad H^{(2)}_1(-1)^{j k} a^{\dagger}_{j+1}\cdots a^{\dagger}_{i_k}\underbrace{a^{\dagger}_{N}}_{k-j+1\text{-th position}}a^{\dagger}_{i_1}\cdots a^{\dagger}_{j-1}|\Omega\rangle
\end{eqnarray}
We should have two terms coming from the above
\begin{eqnarray}
a^{\dagger}_{i_1}\cdots \underbrace{A^{\dagger}}_{j\text{-th position}}\cdots a^{\dagger}_{i_k}|\Omega\rangle\quad\text{and}\quad (-1)^{j k}a^{\dagger}_{j+1}\cdots a^{\dagger}_{i_k}\underbrace{A^{\dagger}}_{k-j+1\text{-th position}}a^{\dagger}_{i_1}\cdots a^{\dagger}_{j-1}|\Omega\rangle
\end{eqnarray}
They have same index content. Rearrangement of the indices in the second of these yields
\begin{eqnarray}
&&(-1)^{(k-j+1)(j-1)+(k-j)+ j k}a^{\dagger}_{i_1}\cdots \underbrace{A^{\dagger}}_{j\text{-th position}}\cdots a^{\dagger}_{i_k}|\Omega\rangle \nonumber \\
&= &(-1)^{-j^2+2 j k+j-1}a^{\dagger}_{i_1}\cdots \underbrace{A^{\dagger}}_{j\text{-th position}}\cdots a^{\dagger}_{i_k}|\Omega\rangle\nonumber \\
&= &(-1)^{-j^2+j}(-1)^{2 j k-1}a^{\dagger}_{i_1}\cdots \underbrace{A^{\dagger}}_{j\text{-th position}}\cdots a^{\dagger}_{i_k}|\Omega\rangle\nonumber \\
&=&-a^{\dagger}_{i_1}\cdots \underbrace{A^{\dagger}}_{j\text{-th position}}\cdots a^{\dagger}_{i_k}|\Omega\rangle
\end{eqnarray}
Thus these terms always cancel and the states 
\begin{eqnarray}
|i_1,\cdots, i_k,N;k\rangle
\end{eqnarray}
are eigenstates of $H^{(2)}_1$:
\begin{eqnarray}
H_1^{(2)}|i_1,\cdots, i_k,N;k\rangle= k\left(^NC_2-N\right)|i_1,\cdots, i_k,N;k\rangle
\end{eqnarray}

\section{Time evolution operator in $1$-particle sector}
\label{app:timeevolution1-fermionsector}
The one-particle Hilbert space spanned by $\{a^{\dagger}_i|\Omega\rangle ;i=1,2,\cdots,N\}$ can equivalently be described by the space spanned by $\{|i\rangle =(0,\cdots ,\underbrace{1}_{i\text{-th position}},\cdots ,0)^T\}$. In this framework, the bilinear Hamiltonian in \eqref{eq:2-cycleH-alltoall} is given by 
\begin{eqnarray}\label{eq:Halltoall-1fermion}
    H:\left(H\right)_{ij}=1-\delta_{ij}\implies H=\mathcal{I}-\mathbb{I}
\end{eqnarray}
where the matrix $\mathcal{I}$ has entries $\mathcal{I}_{ij}=1$ and the matrix $\mathbb{I}$ is the $N\times N$ identity matrix with entries $\mathbb{I}_{ij}=\delta_{ij}$. They satisfy the following properties
\begin{eqnarray}
  \mathcal{I}^m=N^{m-1}\mathcal{I}\quad\text{and}\quad \mathbb{I}^m=\mathbb{I}  
\end{eqnarray}
Then one can simplify : 
\begin{eqnarray}
    e^{-i H t}=\left[\left(\mathbb{I}-\frac{\mathcal{I}}{N}\right)+\frac{\mathcal{I}}{N}e^{-i N t}\right]e^{i t}
\end{eqnarray}
The matrix amplitudes turn out to be
\begin{eqnarray}
    \langle i |e^{-i H t}| j \rangle=\left[\left(\delta_{ij}-\frac{1}{N}\right)+\frac{1}{N}e^{-i N t}\right]e^{it}
\end{eqnarray}
Further, the probability can be calculated as 
\begin{eqnarray}
    |\langle j|j(t)\rangle|^2 &=& \left|\left[\left(1-\frac{1}{N}\right)+\frac{1}{N}e^{-i N t}\right]e^{i t}\right|^2\nonumber \\
    &=&\frac{1}{N^2}\left[1+(N-1)^2+2(N-1)\cos{(Nt)}\right]\\
    |\langle j'|j(t)\rangle|_{\j'\neq j}^2 &=& \left|\left[-\frac{1}{N}+\frac{1}{N}e^{-i N t}\right]e^{i t}\right|^2\nonumber \\
    &=&\frac{2}{N^2}\left[1-\cos{(Nt)}\right]
\end{eqnarray}


\bibliographystyle{ieeetr}
\normalem
\bibliography{refs}

\begin{thebibliography}{10}

\bibitem{1958PhRv..109.1492A}
P.~W. {Anderson}, ``{Absence of Diffusion in Certain Random Lattices},'' {\em
  Physical Review}, vol.~109, pp.~1492--1505, Mar. 1958.

\bibitem{Basko2005MetalinsulatorTI}
D.~M. Basko, I.~Aleiner, and B.~L. Altshuler, ``Metal–insulator transition in
  a weakly interacting many-electron system with localized single-particle
  states,'' {\em Annals of Physics}, vol.~321, pp.~1126--1205, 2005.

\bibitem{PhysRevE.50.888}
M.~Srednicki, ``Chaos and quantum thermalization,'' {\em Phys. Rev. E},
  vol.~50, pp.~888--901, Aug 1994.

\bibitem{PhysRevA.43.2046}
J.~M. Deutsch, ``Quantum statistical mechanics in a closed system,'' {\em Phys.
  Rev. A}, vol.~43, pp.~2046--2049, Feb 1991.

\bibitem{Deutsch_2018}
J.~M. Deutsch, ``Eigenstate thermalization hypothesis,'' {\em Reports on
  Progress in Physics}, vol.~81, p.~082001, jul 2018.

\bibitem{Buvca2023UnifiedTO}
B.~Buvca, ``Unified theory of local quantum many-body dynamics: Eigenoperator
  thermalization theorems,'' 2023.

\bibitem{PhysRevLett.118.266601}
A.~Smith, J.~Knolle, D.~L. Kovrizhin, and R.~Moessner, ``Disorder-free
  localization,'' {\em Phys. Rev. Lett.}, vol.~118, p.~266601, Jun 2017.

\bibitem{PhysRevLett.119.176601}
A.~Smith, J.~Knolle, R.~Moessner, and D.~L. Kovrizhin, ``Absence of ergodicity
  without quenched disorder: From quantum disentangled liquids to many-body
  localization,'' {\em Phys. Rev. Lett.}, vol.~119, p.~176601, Oct 2017.

\bibitem{PhysRevLett.120.030601}
M.~Brenes, M.~Dalmonte, M.~Heyl, and A.~Scardicchio, ``Many-body localization
  dynamics from gauge invariance,'' {\em Phys. Rev. Lett.}, vol.~120,
  p.~030601, Jan 2018.

\bibitem{PhysRevB.102.165132}
I.~Papaefstathiou, A.~Smith, and J.~Knolle, ``Disorder-free localization in a
  simple $u(1)$ lattice gauge theory,'' {\em Phys. Rev. B}, vol.~102,
  p.~165132, Oct 2020.

\bibitem{PhysRevLett.122.040606}
M.~Schulz, C.~A. Hooley, R.~Moessner, and F.~Pollmann, ``Stark many-body
  localization,'' {\em Phys. Rev. Lett.}, vol.~122, p.~040606, Jan 2019.

\bibitem{Morong2021PublisherCO}
W.~Morong, F.~Liu, P.~Becker, K.~S. Collins, L.~Feng, A.~Kyprianidis,
  G.~Pagano, T.~You, A.~V. Gorshkov, and C.~R. Monroe, ``Publisher correction:
  Observation of stark many-body localization without disorder,'' {\em Nature},
  vol.~601, pp.~E13 -- E13, 2021.

\bibitem{PhysRevB.106.174305}
H.~Lang, P.~Hauke, J.~Knolle, F.~Grusdt, and J.~C. Halimeh, ``Disorder-free
  localization with stark gauge protection,'' {\em Phys. Rev. B}, vol.~106,
  p.~174305, Nov 2022.

\bibitem{Wang2021StarkML}
Y.-Y. Wang, Z.-H. Sun, and H.~Fan, ``Stark many-body localization transitions
  in superconducting circuits,'' {\em Physical Review B}, 2021.

\bibitem{Gao2023NonthermalDI}
C.~Gao, Z.~Tang, F.~Zhu, Y.~Zhang, H.~Pu, and L.~Chen, ``Non-thermal dynamics
  in a spin-1/2 lattice schwinger model,'' 2023.

\bibitem{PhysRevResearch.3.L032069}
G.-Y. Zhu and M.~Heyl, ``Subdiffusive dynamics and critical quantum
  correlations in a disorder-free localized kitaev honeycomb model out of
  equilibrium,'' {\em Phys. Rev. Res.}, vol.~3, p.~L032069, Sep 2021.

\bibitem{Wadleigh2022InteractingSL}
L.~Wadleigh, N.~G. Kowalski, and B.~Demarco, ``Interacting stark localization
  dynamics in a three-dimensional lattice bose gas,'' {\em Physical Review A},
  2022.

\bibitem{Zhang2022StableIA}
N.~Zhang, Y.~Ke, L.~Lin, L.~Zhang, and C.~Lee, ``Stable interaction-induced
  anderson-like localization embedded in standing waves,'' {\em New Journal of
  Physics}, vol.~25, 2022.

\bibitem{Peres1984StabilityOQ}
A.~Peres, ``Stability of quantum motion in chaotic and regular systems,'' {\em
  Physical Review A}, vol.~30, pp.~1610--1615, 1984.

\bibitem{Goussev2012LoschmidtE}
A.~Goussev, R.~Jalabert, H.~M. Pastawski, and D.~A. Wisniacki, ``Loschmidt
  echo,'' {\em Scholarpedia}, vol.~7, p.~11687, 2012.

\bibitem{Abanin2018ColloquiumM}
D.~A. Abanin, E.~Altman, I.~Bloch, and M.~Serbyn, ``Colloquium : Many-body
  localization, thermalization, and entanglement,'' {\em Reviews of Modern
  Physics}, 2018.

\bibitem{KitaevSeminar}
A.~Kitaev, ``A simple model of quantum holography,'' {\em KITP strings seminar
  and Entanglement 2015 program},
  vol.~http://online.kitp.ucsb.edu/online/entangled15/., Feb. 12, April 7, and
  May 27, 2015.

\bibitem{Sachdev1992GaplessSG}
Sachdev and Ye, ``Gapless spin-fluid ground state in a random quantum
  heisenberg magnet.,'' {\em Physical review letters}, vol.~70 21,
  pp.~3339--3342, 1992.

\bibitem{Maldacena2015ABO}
J.~Maldacena, S.~H. Shenker, and D.~Stanford, ``A bound on chaos,'' {\em
  Journal of High Energy Physics}, vol.~2016, pp.~1--17, 2015.

\bibitem{Rosenhaus2018AnIT}
V.~Rosenhaus, ``An introduction to the syk model,'' {\em Journal of Physics A:
  Mathematical and Theoretical}, vol.~52, 2018.

\bibitem{Kempe2003QuantumRW}
J.~Kempe, ``Quantum random walks: An introductory overview,'' {\em Contemporary
  Physics}, vol.~44, pp.~307 -- 327, 2003.

\bibitem{VenegasAndraca2012QuantumWA}
S.~E. Venegas-Andraca, ``Quantum walks: a comprehensive review,'' {\em Quantum
  Information Processing}, vol.~11, pp.~1015--1106, 2012.

\bibitem{Reitzner2012QuantumW}
D.~Reitzner, D.~Nagaj, and V.~R. Buzek, ``Quantum walks,'' 2012.

\bibitem{Childs2001AnEO}
A.~M. Childs, E.~Farhi, and S.~Gutmann, ``An example of the difference between
  quantum and classical random walks,'' {\em Quantum Information Processing},
  vol.~1, pp.~35--43, 2001.

\bibitem{Ambainis2001OnedimensionalQW}
A.~Ambainis, E.~Bach, A.~Nayak, A.~Vishwanath, and J.~Watrous,
  ``One-dimensional quantum walks,'' in {\em Symposium on the Theory of
  Computing}, 2001.

\bibitem{aharonov1993quantum}
Y.~Aharonov, L.~Davidovich, and N.~Zagury, ``Quantum random walks,'' {\em
  Physical Review A}, vol.~48, no.~2, p.~1687, 1993.

\bibitem{Santha2008QuantumWB}
M.~Santha, ``Quantum walk based search algorithms,'' in {\em Theory and
  Applications of Models of Computation}, 2008.

\bibitem{Shenvi2002QuantumRS}
N.~Shenvi, J.~Kempe, and K.~B. Whaley, ``Quantum random-walk search
  algorithm,'' {\em Physical Review A}, vol.~67, p.~052307, 2002.

\bibitem{10.5555/2462630}
R.~Portugal, {\em Quantum Walks and Search Algorithms}.
\newblock Springer Publishing Company, Incorporated, 2013.

\bibitem{Bepari2021QuantumWA}
K.~Bepari, S.~Malik, M.~Spannowsky, and S.~Williams, ``Quantum walk approach to
  simulating parton showers,'' {\em Physical Review D}, 2021.

\bibitem{Nachman2019QuantumAF}
B.~P. Nachman, D.~Provasoli, W.~A. de~Jong, and C.~W. Bauer, ``Quantum
  algorithm for high energy physics simulations.,'' {\em Physical review
  letters}, vol.~126 6, p.~062001, 2019.

\bibitem{Somma2008QuantumSO}
R.~D. Somma, S.~Boixo, H.~Barnum, and E.~Knill, ``Quantum simulations of
  classical annealing processes.,'' {\em Physical review letters}, vol.~101 13,
  p.~130504, 2008.

\bibitem{Strauch2005RelativisticQW}
F.~W. Strauch, ``Relativistic quantum walks,'' {\em Physical Review A},
  vol.~73, p.~054302, 2005.

\bibitem{Childs2012UniversalCB}
A.~M. Childs, D.~Gosset, and Z.~Webb, ``Universal computation by multiparticle
  quantum walk,'' {\em Science}, vol.~339, pp.~791 -- 794, 2012.

\bibitem{Underwood2010UniversalQC}
M.~S. Underwood and D.~L. Feder, ``Universal quantum computation by
  discontinuous quantum walk,'' {\em Physical Review A}, vol.~82, p.~042304,
  2010.

\bibitem{Asaka2021TwolevelQW}
R.~Asaka, K.~Sakai, and R.~Yahagi, ``Two-level quantum walkers on directed
  graphs. i. universal quantum computing,'' {\em Physical Review A}, 2021.

\bibitem{Childs2008UniversalCB}
A.~M. Childs, ``Universal computation by quantum walk.,'' {\em Physical review
  letters}, vol.~102 18, p.~180501, 2008.

\bibitem{Sansoni2011TwoparticleBQ}
L.~Sansoni, F.~Sciarrino, G.~Vallone, P.~Mataloni, A.~Crespi, R.~Ramponi, and
  R.~Osellame, ``Two-particle bosonic-fermionic quantum walk via integrated
  photonics.,'' {\em Physical review letters}, vol.~108 1, p.~010502, 2011.

\bibitem{Qin2014StatisticsdependentQC}
X.~Qin, Y.~Ke, X.-W. Guan, Z.~Li, N.~Andrei, and C.~Lee, ``Statistics-dependent
  quantum co-walking of two particles in one-dimensional lattices with
  nearest-neighbor interactions,'' {\em Physical Review A}, vol.~90, p.~062301,
  2014.

\bibitem{Giri2020TwoCQ}
M.~K. Giri, S.~Mondal, B.~P. Das, and T.~Mishra, ``Two component quantum walk
  in one-dimensional lattice with hopping imbalance,'' {\em Scientific
  Reports}, vol.~11, 2020.

\bibitem{Melnikov2013QuantumWO}
A.~A. Melnikov and L.~Fedichkin, ``Quantum walks of interacting fermions on a
  cycle graph,'' {\em Scientific Reports}, vol.~6, 2013.

\bibitem{Melnikov2018HittingTF}
A.~A. Melnikov, A.~P. Alodjants, and L.~Fedichkin, ``Hitting time for quantum
  walks of identical particles,'' in {\em International Conference on Micro-
  and Nano-Electronics}, 2018.

\bibitem{Lahini2011QuantumWO}
Y.~Lahini, M.~Verbin, S.~D. Huber, Y.~Bromberg, R.~Pugatch, and Y.~R.
  Silberberg, ``Quantum walk of two interacting bosons,'' {\em Physical Review
  A}, vol.~86, p.~011603, 2011.

\bibitem{Wiater2017TwoBQ}
D.~Wiater, T.~Sowi'nski, and J.~J. Zakrzewski, ``Two bosonic quantum walkers in
  one-dimensional optical lattices,'' {\em Physical Review A}, vol.~96,
  p.~043629, 2017.

\bibitem{Krovi2007SymmetryIQ}
H.~Krovi, ``Symmetry in quantum walks,'' 2007.

\bibitem{Janmark2014GlobalSI}
J.~Janmark, D.~A. Meyer, and T.~G. Wong, ``Global symmetry is unnecessary for
  fast quantum search,'' {\em Physical Review Letters}, vol.~112, p.~210502,
  2014.

\bibitem{PhysRevA.76.022316}
C.~M. Chandrashekar, R.~Srikanth, and S.~Banerjee, ``Symmetries and noise in
  quantum walk,'' {\em Phys. Rev. A}, vol.~76, p.~022316, Aug 2007.

\bibitem{Cedzich2016TheTC}
C.~Cedzich, T.~Geib, F.~A. Gr{\"u}nbaum, C.~Stahl, L.~Vel{\'a}zquez, A.~H.
  Werner, and R.~F. Werner, ``The topological classification of one-dimensional
  symmetric quantum walks,'' {\em Annales Henri Poincar{\'e}}, vol.~19,
  pp.~325--383, 2016.

\bibitem{Cedzich2019QuantumWS}
C.~Cedzich, T.~Geib, F.~A. Gr{\"u}nbaum, L.~Vel'azquez, A.~H. Werner, and R.~F.
  Werner, ``Quantum walks: Schur functions meet symmetry protected topological
  phases,'' {\em Communications in Mathematical Physics}, vol.~389, pp.~31 --
  74, 2019.

\bibitem{Geib2019TopologicalAO}
T.~Geib, C.~Cedzich, A.~H. Werner, and R.~F. Werner, ``Topological aspects of
  discrete and continuous time quantum walks on one dimensional lattices,''
  2019.

\bibitem{Cedzich2018CompleteHI}
C.~Cedzich, T.~Geib, C.~Stahl, L.~Vel{\'a}zquez, A.~H. Werner, and R.~F.
  Werner, ``Complete homotopy invariants for translation invariant symmetric
  quantum walks on a chain,'' {\em arXiv: Quantum Physics}, 2018.

\bibitem{Danac2020DisorderfreeLI}
B.~Danacı, I.~Yalçinkaya, B.~Çakmak, G.~Karpat, S.~P. Kelly, and A.~L.
  Subaşı, ``Disorder-free localization in quantum walks,'' {\em arXiv:
  Quantum Physics}, 2020.

\bibitem{Mandal2021LocalizationOT}
A.~Mandal, R.~S. Sarkar, and B.~Adhikari, ``Localization of two dimensional
  quantum walks defined by generalized grover coins,'' {\em Journal of Physics
  A: Mathematical and Theoretical}, vol.~56, 2021.

\bibitem{Singh2017InterferenceAC}
S.~Singh and C.~M. Chandrashekar, ``Interference and correlated coherence in
  disordered and localized quantum walk,'' {\em arXiv: Quantum Physics}, 2017.

\bibitem{Joye2012DynamicalLF}
A.~Joye, ``Dynamical localization for d-dimensional random quantum walks,''
  {\em Quantum Information Processing}, vol.~11, pp.~1251--1269, 2012.

\bibitem{Cedzich2019AndersonLF}
C.~Cedzich and A.~H. Werner, ``Anderson localization for electric quantum walks
  and skew-shift cmv matrices,'' {\em Communications in Mathematical Physics},
  vol.~387, pp.~1257 -- 1279, 2019.

\bibitem{Gerhardt2003ContinuousTimeQW}
H.~Gerhardt and J.~Watrous, ``Continuous-time quantum walks on the symmetric
  group,'' in {\em RANDOM-APPROX}, 2003.

\bibitem{PhysRevE.55.4067}
E.~Br\'ezin and S.~Hikami, ``Spectral form factor in a random matrix theory,''
  {\em Phys. Rev. E}, vol.~55, pp.~4067--4083, Apr 1997.

\bibitem{cotler2017black}
J.~S. Cotler, G.~Gur-Ari, M.~Hanada, J.~Polchinski, P.~Saad, S.~H. Shenker,
  D.~Stanford, A.~Streicher, and M.~Tezuka, ``Black holes and random
  matrices,'' {\em Journal of High Energy Physics}, vol.~2017, no.~5,
  pp.~1--54, 2017.

\bibitem{cipolloni2023spectral}
G.~Cipolloni, L.~Erd{\H{o}}s, and D.~Schr{\"o}der, ``On the spectral form
  factor for random matrices,'' {\em Communications in Mathematical Physics},
  pp.~1--36, 2023.

\bibitem{Arenz2013ControlOO}
C.~Arenz, G.~Gualdi, and D.~Burgarth, ``Control of open quantum systems: case
  study of the central spin model,'' {\em New Journal of Physics}, vol.~16,
  2013.

\bibitem{Nepomechie2018TheSH}
R.~I. Nepomechie and X.-W. Guan, ``The spin-s homogeneous central spin model:
  exact spectrum and dynamics,'' {\em Journal of Statistical Mechanics: Theory
  and Experiment}, vol.~2018, 2018.

\bibitem{Villazon2020PersistentDS}
T.~Villazon, P.~W. Claeys, M.~Pandey, A.~Polkovnikov, and A.~Chandran,
  ``Persistent dark states in anisotropic central spin models,'' {\em
  Scientific Reports}, vol.~10, 2020.

\bibitem{Godsil2017SedentaryQW}
C.~D. Godsil, ``Sedentary quantum walks,'' {\em Linear Algebra and its
  Applications}, vol.~614, pp.~356--375, 2017.

\bibitem{Carlson2006UniversalMO}
W.~Carlson, A.~For, E.~Harris, J.~Rosen, C.~Tamon, and K.~Wrobel, ``Universal
  mixing of quantum walk on graphs,'' {\em Quantum Inf. Comput.}, vol.~7,
  pp.~738--751, 2006.

\bibitem{Xu2009ExactAR}
X.~Xu, ``Exact analytical results for quantum walks on star graphs,'' {\em
  Journal of Physics A: Mathematical and Theoretical}, vol.~42, p.~115205,
  2009.

\bibitem{Frigerio2022SwiftCQ}
M.~Frigerio and M.~G. Paris, ``Swift chiral quantum walks,'' {\em Linear
  Algebra and its Applications}, 2022.

\bibitem{Godsil2010WhenCP}
C.~D. Godsil, ``When can perfect state transfer occur,'' {\em arXiv:
  Combinatorics}, 2010.

\bibitem{Ide2014LocalSS}
Y.~Ide, ``Local subgraph structure can cause localization in continuous-time
  quantum walk,'' {\em arXiv: Quantum Physics}, 2014.

\bibitem{Godsil2001AlgebraicGT}
C.~D. Godsil and G.~F. Royle, ``Algebraic graph theory,'' in {\em Graduate
  texts in mathematics}, 2001.

\bibitem{Brouwer2011SpectraOG}
A.~E. Brouwer and W.~H. Haemers, ``Spectra of graphs,'' 2011.

\bibitem{tsomokos2008fully}
D.~I. Tsomokos, S.~Ashhab, and F.~Nori, ``Fully connected network of
  superconducting qubits in a cavity,'' {\em New Journal of Physics}, vol.~10,
  no.~11, p.~113020, 2008.

\bibitem{Xu2019ProbingTD}
K.~Xu, Z.-H. Sun, W.~Liu, Y.-R. Zhang, H.~Li, H.~Dong, W.~Ren, P.~Zhang,
  F.~Nori, D.~Zheng, H.~Fan, and H.~Wang, ``Probing the dynamical phase
  transition with a superconducting quantum simulator,'' {\em arXiv: Quantum
  Physics}, 2019.

\bibitem{Song2019GenerationOM}
C.~Song, K.~Xu, H.~Li, Y.-R. Zhang, X.~Zhang, W.~Liu, Q.~Guo, Z.~Wang, W.~Ren,
  J.~Hao, H.~Feng, H.~Fan, D.~Zheng, D.-W. Wang, H.~Wang, and S.-Y. Zhu,
  ``Generation of multicomponent atomic schr{\"o}dinger cat states of up to 20
  qubits,'' {\em Science}, vol.~365, pp.~574 -- 577, 2019.

\bibitem{Hazra2020LongrangeCI}
S.~Hazra, A.~Bhattacharjee, M.~Chand, K.~V. Salunkhe, S.~Gopalakrishnan, M.~P.
  Patankar, and R.~Vijay, ``Long-range connectivity in a superconducting
  quantum processor using a ring resonator.,'' {\em arXiv: Quantum Physics},
  2020.

\bibitem{Lanyon2011UniversalDQ}
B.~P. Lanyon, C.~Hempel, D.~Nigg, M.~M{\"u}ller, R.~Gerritsma,
  F.~Z{\"a}hringer, P.~Schindler, J.~T. Barreiro, M.~Rambach, G.~Kirchmair,
  M.~Hennrich, P.~Zoller, R.~Blatt, and C.~F. Roos, ``Universal digital quantum
  simulation with trapped ions,'' {\em Science}, vol.~334, pp.~57 -- 61, 2011.

\bibitem{Wright2019BenchmarkingA1}
K.~Wright, K.~M. Beck, S.~Debnath, J.~M. Amini, Y.~S. Nam, N.~Grzesiak, J.-S.
  Chen, N.~C. Pisenti, M.~Chmielewski, C.~Collins, K.~M. Hudek, J.~Mizrahi,
  J.~D. Wong-Campos, S.~Allen, J.~Apisdorf, P.~Solomon, M.~Williams, A.~M.
  Ducore, A.~Blinov, S.~M. Kreikemeier, V.~Chaplin, M.~J. Keesan, C.~Monroe,
  and J.~Kim, ``Benchmarking an 11-qubit quantum computer,'' {\em Nature
  Communications}, vol.~10, 2019.

\bibitem{Cazalilla2014UltracoldFG}
M.~A. Cazalilla and A.~M. Rey, ``Ultracold fermi gases with emergent su(n)
  symmetry,'' {\em Reports on Progress in Physics}, vol.~77, 2014.

\bibitem{Hashimoto2017OutoftimeorderCI}
K.~Hashimoto, K.~Murata, and R.~Yoshii, ``Out-of-time-order correlators in
  quantum mechanics,'' {\em Journal of High Energy Physics}, vol.~2017,
  pp.~1--31, 2017.

\bibitem{Balachandran2013EntanglementAP}
A.~P. Balachandran, T.~R. Govindarajan, A.~R. de~Queiroz, and A.~F. Reyes-Lega,
  ``Entanglement and particle identity: a unifying approach.,'' {\em Physical
  review letters}, vol.~110 8, p.~080503, 2013.

\bibitem{Balachandran2013AlgebraicAT}
A.~P. Balachandran, T.~R. Govindarajan, A.~R. Queiroz, and A.~F. Reyes-Lega,
  ``Algebraic approach to entanglement and entropy,'' {\em Physical Review A},
  vol.~88, p.~022301, 2013.

\bibitem{Reyes-Lega:2022sbq}
A.~F. Reyes-Lega, ``{Entanglement Entropy in Quantum Mechanics: An Algebraic
  Approach},'' 12 2022.

\bibitem{Balachandran:2012pa}
A.~P. Balachandran, A.~R. de~Queiroz, and S.~Vaidya, ``{Entropy of Quantum
  States: Ambiguities},'' {\em Eur. Phys. J. Plus}, vol.~128, p.~112, 2013.

\bibitem{Balachandran:2013kia}
A.~P. Balachandran, A.~R. de~Queiroz, and S.~Vaidya, ``{Quantum Entropic
  Ambiguities: Ethylene},'' {\em Phys. Rev. D}, vol.~88, no.~2, p.~025001,
  2013.

\bibitem{Benatti2020EntanglementII}
F.~Benatti, R.~Floreanini, F.~Franchini, and U.~Marzolino, ``Entanglement in
  indistinguishable particle systems,'' {\em Physics Reports}, 2020.

\end{thebibliography}



\end{document}